\documentclass[10pt]{article}
\usepackage{amsmath,amssymb,graphicx} 
\usepackage{epsf}
\usepackage{pstricks}
\usepackage{cite}
\usepackage[bookmarks]{hyperref}
\usepackage{booktabs}

\newcommand{\beq}{\begin{eqnarray}}
\newcommand{\eeq}{\end{eqnarray}}

\newcommand{\centeron}[2]{{\setbox0=\hbox{#1}\setbox1=\hbox{#2}\ifdim
                                        
\wd1>\wd0\kern.5\wd1\kern-.5\wd0\fi
\copy0

\kern-.5\wd0\kern-.5\wd1\copy1\ifdim\wd0>\wd1
                                       \kern.5\wd0\kern-.5\wd1\fi}}
\newcommand{\ltap}{\>\centeron{\raise.35ex\hbox{$<$}}
                               {\lower.65ex\hbox{$\sim$}}\>}
\newcommand{\gtap}{\>\centeron{\raise.35ex\hbox{$>$}}
                               {\lower.65ex\hbox{$\sim$}}\>}

\newcommand\ZZ{\hbox{\zfont Z\kern-.4emZ}}
\font\zfont = cmss10 

\numberwithin{equation}{section}

\begin{document}
\begin{titlepage}
\begin{flushright}
CP3-13-22, LYCEN 2013-03, SHEP-13-10
\end{flushright}

\vskip.5cm
\begin{center}
{\huge \bf 
Model Independent Framework for Searches of Top Partners
}

\vskip.1cm
\end{center}
\vskip0.2cm

\begin{center}
{\bf
{Mathieu Buchkremer}$^{a}$, {Giacomo Cacciapaglia}$^{b,c}$, {Aldo Deandrea}$^{b}$, {Luca Panizzi}$^{c}$}
\end{center}
\vskip 8pt

\begin{center}
{\small
$^a$ {\it Centre for Cosmology, Particle Physics and Phenomenology (CP3), \\
Universit\'e catholique de Louvain, Chemin du Cyclotron, 2, B-1348, Louvain-la-Neuve, Belgium}\\
\vspace*{0.1cm}
$^{b}$ {\it Universit\'e de Lyon, F-69622 Lyon, France; Universit\'e Lyon 1, Villeurbanne;
CNRS/IN2P3, \\UMR5822, Institut de Physique Nucl\'eaire de Lyon, 
F-69622 Villeurbanne Cedex, France } \\
\vspace*{0.1cm}
$^c${\it School of Physics and Astronomy, University of Southampton,\\ Highfield, Southampton SO17 1BJ, UK}
}
\end{center}

\vglue 0.3truecm

\begin{abstract}
\vskip 3pt
\noindent
We propose a model-independent and general framework to study the LHC phenomenology of top partners, i.e. Vector-Like quarks 
including particles with different electro-magnetic charge.
We consider Vector-Like quarks embedded in general representations of the weak $SU(2)_{L}$, coupling to all Standard Model quarks 
via Yukawa mixing focusing on the case of a single multiplet. We show that, with very minimal and quite general assumptions, top partners may be studied in terms of few 
parameters in an effective Lagrangian description with a clear and simple connection with experimental observables. 
We also demonstrate that the parametrisation can be applied as well to cases with many Vector-like multiplets, thus covering most realistic models of New Physics.
We perform a numerical study to understand the conclusions which can be drawn within such a description and the expected potential 
for discovery or exclusion at the LHC. Our main results are a clear connection between branching ratios and single production 
channels, and the identification of novel interesting channels to be studied at the LHC. 

\end{abstract}
\end{titlepage}

\newpage

\tableofcontents


\section{Introduction}
\label{sec:intro}
\setcounter{equation}{0}
\setcounter{footnote}{0}

After the discovery of a new resonance that matches the expectations for a Standard Model Higgs boson, the LHC experiments will 
now focus on New Physics Searches. One of the Holy Grails of particle physics is the issue of naturalness, and of course the solution 
of the mystery of the nature of Dark Matter, which may be produced at the LHC and would appear as missing transverse momentum.
The discovery of a Higgs candidate at a mass of 125 GeV, thus very close to the electroweak scale, is in fact a realisation of the 
naturalness problem: why is the scalar mass so close to the electroweak scale? What symmetry, if any, is shielding it from large loop 
corrections from heavy physics? 

The quest for an answer to such questions has been the guiding principle behind the flourishing of model building in the past decades.
A general assumption of these models is the presence of new weakly coupled states which effectively cut-off the divergent loop 
contributions to the Higgs mass from Standard Model states, mainly the top quark and the massive gauge bosons $W^\pm$ and $Z$.
The absence of fine-tuning would therefore require the masses of the hypothetical new states to lay below or around the TeV scale.
As the top quark is known to have the largest Yukawa coupling to the Higgs field, it is a natural expectation that the lightest new states 
are partners of the top itself.
Supersymmetry, the most popular theory of New Physics, predicts the existence of states with different statistics: therefore, the top 
quark would be complemented by scalar tops. Direct searches for supersymmetry have by now pushed the limit on the masses of 
supersymmetric states in simplified models well above the TeV threshold~(see for example 
\cite{Chatrchyan:2013lya,Chatrchyan:2011ek,Aad:2012naa,Aad:2012fqa} for recent analyses). However, this applies to partners of the 
gluons and light quarks, which are abundantly produced at the LHC but play a marginal role in the naturalness argument.
Direct limits on the top superpartners~\cite{Aad:2012uu,CMS-PAS-SUS-13-011}, and the partners of the $W$ and $Z$ (charginos and 
neutralinos) \cite{Aad:2012hba,Chatrchyan:2012pka}, on the other hand, are still well below the TeV scale due to the smaller 
production rates, and the more challenging final states from the experimental point of view. The naturalness argument, therefore, is far 
from being in crisis! The main conclusion to be drawn is that indeed we are entering an era in which these ideas can start to be tested 
with the LHC.

From the theoretical side there are several models beyond supersymmetry which also address the naturalness issue or postpone it 
to higher scales. Some are based on the effective Lagrangian approach, others introduce extended global symmetries 
(Little Higgs models)~\cite{littlehiggs}, extra dimensional space symmetries (Gauge-Higgs 
Unification)~\cite{Antoniadis:2001cv,Hosotani:2004wv,Agashe:2004rs} or assume that the breaking of the 
electroweak symmetry is due to a strongly interacting dynamics (Composite Higgs models)~\cite{Agashe:2004rs}.
Modern incarnations of Technicolour, which have a light Higgs-like scalar in the spectrum~\cite{Matsuzaki:2012mk,Foadi:2012bb}, 
should also be included in the list. In all the above cases, a common prediction is the presence of partners of the top quark and more 
generally multiplets containing a top partner of the vector-like type\cite{delAguila:1982fs}, which have 
the same spin and only differ in the embedding into representations of the weak isospin, $SU(2)_{L}$.
They typically arise as Kaluza-Klein recursions of the quarks in models of extra dimensions~\cite{Contino}, states needed to complete 
a full representation of the extended symmetries or additional massive composite states of the strong 
dynamics~\cite{Contino,Matsedonskyi:2012ym,Dissertori:2010ug}. Also, the possibility for new heavy quarks featuring 
$s$-channel resonances remains of prime interest at the LHC. 
Contrary to sequential fourth family quarks which are heavily constrained from the Higgs boson searches due to their non-decoupling properties \cite{Eberhardt:2012gv}, indirect bounds on 
non-chiral quarks are much weaker: they nevertheless affect the properties of the Higgs \cite{Moreau:2012da,Bonne:2012im}, 
for instance affecting the production of a pair of Higgses~\cite{Contino:2012xk}, or offering new Higgs production 
mechanisms~\cite{AguilarSaavedra:2006gw,Azatov:2012rj,Harigaya:2012ir,Carmona:2012jk}. The phenomenology of new heavy quarks has been widely 
studied in literature, see for 
example~\cite{Lavoura:1992np,delAguila:2000rc,AguilarSaavedra:2005pv,Cynolter:2008ea,AguilarSaavedra:2009es,Berger:2012ec,Okada:2012gy} 
and the forthcoming direct searches at the LHC will therefore play a fundamental role in testing the large number of models 
predicting the existence of these states.

The importance of top partners, or generically new quarks, is also supported by the massive ongoing experimental effort for their discovery: many searches are 
being done by both CMS and ATLAS.
The present limits on their masses have reached scales around 700 GeV, reaching the mass range of interest for the 
naturalness argument. The first searches performed on the 2011 dataset at 7 TeV of centre-of-mass energy, as it is often the case, 
had to rely on simplifying assumptions: typically, the hypothesis of 100\% decays into a single channel. 
Such assumption is unrealistic, in the sense that models tend to predict rates into various channels, however this simplification allowed 
for manageable interpretation of the data and exploration of the reach of the experiments.
Recently, more complete searches have been performed, a review of which can be found in \cite{Okada:2012gy,Garberson:2013jz}. 
For instance, ATLAS published a search for a top partner with charge 
$2/3$, $t'$, assuming decays into three channels, $W^+ b$, $Z t$ and $H t$, and scanning over various combinations of the branching 
ratios \cite{ATLAS:2013ima}. Although searches for new heavy quarks decaying via $Q\rightarrow Wq$ where $%
q=u,d,c,s,b$ for up- and down-like $Q$ quarks have been performed \cite{Aad:2011yn,Aad:2012bt}, only decays into the third generation quarks have been considered in the reinterpretations so far. 

At present most of the experimental searches assume that the new heavy quarks are QCD pair produced, while combining searches 
for quarks which are either produced singly or pairwise (also via EW interactions) will become a more effective option in the near future. Indeed, present limits from the LHC start to enter the region in which single production becomes relevant and this will be more and more the case if the bounds are raised. Experimental searches, such as e.g. \cite{Aad:2013rna}, now focus on the electroweak single production of new heavy quarks, through channels that are sensitive to the size of the coupling to the 
standard quarks. However, most of the current studies are based on the assumption that the new quarks only couple to the first generation of quarks, and final states involving tops and/or bottoms have not been explored. We have therefore at present an incomplete picture, which will be for sure studied in more detail in the near future using more general analyses and a larger amount of data. 
While many decay channels and final states have been already considered by the experimental teams, little attempt has been done to 
combine the information extracted from the data in a systematic way: the main reason for this is the lack of a complete and model-independent framework to describe the interactions of such new particles.
In Ref.~\cite{Atre:2011ae} a general Lagrangian for the interactions of new Vector-Like (VL) quarks with the first generation of quarks has been 
proposed, and completed in Ref.~\cite{Atre:2013ap} to include the interactions with a Higgs, and used to study single production at the 
LHC. However, couplings with the other quark families are missing: in particular, decays to the tops and/or bottoms would change the 
search strategies and therefore significantly affect the limits.
In Ref.~\cite{DeSimone:2012fs}, a ``hunter guide'' for top partners has been proposed: the effective models proposed there are heavily relying on the assumption that they belong to a model of composite Higgs. For 
instance, only couplings to third generation quarks, which are assumed to be more composite that light ones, are included, and the effects of off-diagonal $Z$ couplings are not considered, based on symmetry reasons designed to remove large corrections to flavour observables and the $Z$ couplings.
Nevertheless, within this class of models, a truly ``model independent'' approach is used, based on the fact that the events generated by the decays of each new quark can be grouped in classes characterised by a universal and model independent experimental efficiency. Every model can be therefore obtained by combining the efficiencies by the effective cross section of each class of events.

In this paper, we propose to combine the two approaches: identify a minimal Lagrangian which describes all the allowed couplings of 
VL quarks, and use it to define model-independent search strategies allowing to fully constrain the masses of top partners in 
any given model.
The main guiding principle is the fact that the decays and single production of the new states are generated via mixing with the standard quarks~\cite{delAguila:2000rc}, induced by Yukawa interactions with the Higgs, thus generalising the ideas discussed in our previous works \cite{Cacciapaglia:2010vn,Cacciapaglia:2011fx}.
This is not an assumption, as any model including VL quarks can be written down as an effective model with this kind of mixing mechanism in place.
The minimal set of parameters we identify, consists of a set mainly describing  the branching ratios into the massive bosons, 
$W^\pm$, $Z$ and the Higgs $H$, and into the three Standard Model quark generations, plus a parameter describing the strength of the 
mixing. The first step would be to study the efficiency of the present searches on final states which have not been considered in the 
simplified assumptions adopted by the collaborations in the first run of the LHC, however without any model-bias 
on the parametrisation. We also propose to use this framework to find corners of the parameter space which are poorly covered 
by present searches, and define new dedicated searches. Finally, the connection between the branching ratios in different channels and 
the single production cross sections can be exploited to extend the searches to include both single and pair production channels, and 
extract a reliable bound on the mixing parameters with the standard quarks. 
This study would allow then to apply the direct search bounds to any model of New Physics which predicts the presence of VL quarks.

In this work we are interested in VL quarks which can mix and decay directly into Standard Model quarks.
Another possibility is to assume that the new quarks are charged under some parity, so that they can only decay into an ordinary quark 
via a new boson which can eventually play the role of Dark Matter~\cite{Meade:2006dw,Alwall:2010jc}: this happens in extra dimensional models with a K-parity~\cite{Appelquist:2000n,Servant:2002aq} or in Little Higgs models with T-parity~\cite{Low:2004xc,Matsumoto:2008fq}. The Dark Matter candidate can be either a spin one (like in 5D models, or in Little Higgs ones) or a spin zero particle (like in 6D models~\cite{Cacciapaglia:2009pa}) state. This scenario will be considered in a future publication~\footnote{Bounds on a fermion decaying into a top plus a stable neutral boson can be extracted in searches based on $t \bar{t}$ plus missing transverse energy.}.

Like any model-independent parametrisation of New Physics, we do rely on minimal and basic assumptions, which are also compatible 
with the principles behind the searches themselves:
\begin{itemize}
\item[-] the new states are embedded in complete representations of $SU(2)_{L}$, and the structure of the mixing is dictated by the Higgs 
field. The specific form of the Higgs sector does not affect our conclusions significantly, as additional doublets will generate the same structures as the SM Higgs, while the vacuum expectation values of other representations are bound to be small due to corrections to the $\rho$ parameter.
\item[-] we study the lightest states of each kind, assuming small mixings with eventual heavier particles, following the interpretation 
of the searches which is based on the presence of a single new state. However, this assumption can be removed and our parametrisation be used to also study the case of many VL quarks with similar masses.
\item[-] the couplings to standard quarks are chiral. This is a direct consequence of the two points above. Remarkably, this property also holds in scenarios with multiple vector-like quark representations. A proof of the previous statement is given in the Appendix~\ref{app:multiVL}, where the consequence of such property will be discussed in more detail.
\end{itemize}
One may also include further biases coming from the limited number of choices for the $SU(2)_{L}$ representations (each representation fixes 
the rates into $W$, $Z$ and $H$), and assumptions coming from the strong bounds from flavour observables (which constrain the rates into different generations).
We will use the latter model-dependent biases to define some benchmark points, but not to limit the validity of the proposed parametrisation.

This minimal set of assumptions makes our present framework consistent with several new physics scenarios. Considering exclusive decays through charged currents, our formalism applies directly to t' (Q$=2/3$) and b' (Q$=-1/3$) chiral quarks. It also remains valid in the context of Composite Higgs models, where top partners are known to arise as light custodians, with masses expected in the range $500-1500$ GeV. After the spontaneous breaking of the electroweak symmetry, it is known that the fermionic spectrum leads in this case to a tower of vector-like resonances of electromagnetic charges $2/3$, -$1/3$ and $5/3$ ~\cite{Contino,Matsedonskyi:2012ym,Dissertori:2010ug}. 
Little Higgs models are also known to allow for several VL quarks, appearing in complete multiplets of the underlying symmetry group. However, most of the corresponding realisations require to introduce new scalars beyond a single Higgs doublet. Unlike the minimal top partners models, new top partners may then decay to Pseudo-Nambu-Goldstone bosons such as, e.g., a second Higgs doublet ~\cite{Kearney:2013oia}. Although top partners can have sizeable branching ratios to such an extended Higgs sector, their decays to non-SM particles are generally expected to be subdominant in most scenarios due to phase space suppression. In our parametrisation, we do cover the case where VL quark decays to new neutral and charged scalars are suppressed.  Under this simplifying assumption, our formalism naturally extends to various classes of models, as long as the lightest states decay mainly to SM quarks via $W$, $Z$ and Higgs bosons. 
The phenomenology of multiple non-degenerate quarks, decaying into the same final states through different channels, will be investigated in a future work. While most of the aforementioned new physics scenarios contain more than one vector-like quark, the direct searches are generally carried out with the prior that only one new state beside SM is present.
Furthermore, the parametrisation presented in this work can be used in the context of multi VL models as long as longer decay chains with VL quarks decaying into each other are considered separately - and, inevitably, in a model-dependent way.

The paper is organised as follows: as a simple preliminary exercise, we discuss in Section~\ref{sec:warmup} the case of a partner of the top quark, i.e. a colour triplet with charge $+2/3$. 
We then generalise the parametrisation, in Section~\ref{sec:param}, to include all the VL quarks that are allowed to decay into standard quarks, thus covering all the quantum numbers present in multiplets that can couple to standard quarks via the Higgs.
The effective Lagrangian is written in terms of the few relevant parameters which have a clear link to observed 
quantities. In Section~\ref{sec:prod} we use this parametrisation to study production and decays of these states, and expand single and pair production cross sections in terms of model-independent coefficients. 
In Section~\ref{sec:analysis}, we perform a 
numerical analysis using a Monte Carlo implementation of the model, and discuss a few outcomes reflecting the expected potential for discovery or exclusion at the LHC. Our conclusions are presented in Section~\ref{sec:conclusion}.

\section{Warm up: the case of a top partner $T$ ($t'$)}
\label{sec:warmup}

For concreteness, we first focus on the case of a top partner $T$, i.e. a VL quark with the same electric charge (and colour) as the top quark.
The most general couplings of a single $T$ with the electroweak gauge bosons can be parametrised as 
\begin{eqnarray}
\mathcal{L}_{T single} &=&  \kappa_W V_{L/R}^{4i} \frac{g}{\sqrt{2}}\; [\bar{T}_{L/R} W_\mu^+ \gamma^\mu d^i_{L/R} ]  + \kappa_Z V_{L/R}^{4i} \frac{g}{2 c_W} \; [\bar{T}_{L/R} Z_\mu \gamma^\mu u^i_{L/R} ]  \nonumber \\
&-& \kappa_H V_{L/R}^{4i} \frac{M}{v}\; [\bar{T}_{R/L} H u^i_{L/R} ] + h.c. \label{eq:topL}
\end{eqnarray}
while the couplings with gluon and photon are standard and dictated by gauge invariance~\footnote{Couplings to the $Z$, and to the $W$ and other VL quarks, are also in general present and they depend on the representation of $SU(2)_{L}$ $T$ belongs to.}. This is a generalisation of the Lagrangian in~\cite{Atre:2011ae} by the inclusion of couplings with the Higgs~\cite{Atre:2013ap}, and to all the generations of 
quarks at the same time. In this formula, $M$ is the mass of the VL quark, $V_{L/R}^{4i}$ represent the mixing matrices between the new quarks 
and the three Standard Model generations labelled by $i$, while the parameters $\kappa_V$ ($V = W$, $Z$, $H$) encode the coupling to the three bosons. The 
normalisation is chosen so that for $\kappa_W =\kappa_Z = \kappa_H=1$, the VL top decays 25\% to $Z$ and $H$ and 50\% to $W$ in the asymptotic limit 
where the mass $M$ goes to infinity, in agreement with what is expected from the Goldstone equivalence theorem. The values of the $\kappa_V$'s are determined by the $SU(2)_{L}$ representation $T$ belongs to, and eventually by mixing to other 
VL representations.

In the most general set-up, $T$ may have sizeable couplings to both left- and right-handed Standard Model quarks $q$. However, in the case of one single light VL quark, 
which is the simple case studied experimentally, it is easy to show that only one of the two mixing angles is large, the other being suppressed by a factor of 
$m_q/M$~\cite{Cacciapaglia:2010vn}. Following this observation, we can simplify the parametrisation by neglecting the suppressed mixing angles, so that the Lagrangian we showed above will only contain one of the two chiral couplings: this approximation may not be precise for the top quark, while it is numerically well justified for all other quarks. 
A discussion of the terms suppressed by the top mass $m_t$, which are generally model-dependent, can be found in Appendix~\ref{app:topmass}.

From the Lagrangian in Eq.(\ref{eq:topL}), the partial widths in the various channels are given by
\beq
\Gamma (T \to W d_i) &=& \kappa_W^2  |V_{L/R}^{4i}|^2 \frac{M^3 g^2}{64 \pi m_W^2} \Gamma_{W} (M, m_W, m_{d_i}) \label{eq:GammaW}\,, \\
\Gamma (T \to Z u_i) &=& \kappa_Z^2 |V_{L/R}^{4i}|^2  \frac{M^3 g^2}{64 \pi m_W^2} \Gamma_{Z} (M, m_Z, m_{u_i})\label{eq:GammaZ}\,, \\
\Gamma (T \to H u_i) &=& \kappa_H^2  |V_{L/R}^{4i}|^2  \frac{M^3 g^2}{64 \pi m_W^2} \Gamma_{H} (M, m_H, m_{u_i})\label{eq:GammaH}\, ,
\eeq
where the kinematic functions are
\beq
\Gamma_{W} &=& \lambda^{\frac{1}{2}} (1, \frac{m_q^2}{M^2}, \frac{m_W^2}{M^2}) \left[ \left( 1-\frac{m_q^2}{M^2} \right)^2 + \frac{m_W^2}{M^2} - 2 \frac{m_W^4}{M^4} + \frac{m_W^2 m_q^2}{M^4} \right]\,, \\
\Gamma_{Z} &=& \frac{1}{2} \lambda^{\frac{1}{2}} (1, \frac{m_q^2}{M^2}, \frac{m_Z^2}{M^2})  \left[ \left( 1-\frac{m_q^2}{M^2} \right)^2 + \frac{m_Z^2}{M^2} - 2 \frac{m_Z^4}{M^4} + \frac{m_Z^2 m_q^2}{M^4} \right]\,, \\
\Gamma_{H} &=& \frac{1}{2} \lambda^{\frac{1}{2}} (1, \frac{m_q^2}{M^2}, \frac{m_H^2}{M^2})  \left[ 1 + \frac{m_q^2}{M^2} -  \frac{m_H^2}{M^2} \right]\,;
\eeq
and the function $\lambda (a, b, c)$ is given by
\beq
\lambda (a, b, c) = a^2 + b^2 + c^2 - 2 ab - 2 ac - 2 bc\,.
\eeq
We expressed the partial width in a fashion that underlines the universal coupling factor, so that the difference between various channels only depends on the masses: for the light quarks, the mass dependence is very mild, therefore we can 
assume that the numbers are the same for all generations. This is not true in general for the top quark, for which the effect of its mass may be important: as we neglected it in the mixing angles, we will consistently neglect it here and comment on its effect at the end of the section and in Appendix \ref{app:topmass}.

Neglecting all quark masses, therefore, the branching ratios can be written as:
\beq
BR(T \to V q_i) = \frac{\kappa_V^2 |V_{L/R}^{4i}|^2 \Gamma_{V}^0}{\left( \sum_{j=1}^3 |V_{L/R}^{4j}|^2 \right) \left(\sum_{V^{\prime }=W,Z,H}
\kappa_{V^{\prime }}^2 \Gamma_{V^{\prime}}^0 \right)}
\eeq
where $\Gamma_{V}^0$ are the kinematic functions for zero quark mass $m_q = 0$:
 \beq
\Gamma_{W}^0 &=& \left( 1 -3  \frac{m_W^4}{M^4} + 2 \frac{m_W^6}{M^6}  \right) \sim 1 + \mathcal{O} (M^{-4})\,, \\
\Gamma_{Z}^0 &=& \frac{1}{2} \left(1 - 3 \frac{m_Z^4}{M^4} + 2 \frac{m_Z^6}{M^6}  \right) \sim \frac{1}{2} + \mathcal{O} (M^{-4})\,, \\
\Gamma_{H}^0 &=& \frac{1}{2}  \left(1- \frac{m_H^2}{M^2}\right)^2 \sim \frac{1}{2} - \frac{m_H^2}{M^2}+ \mathcal{O} (M^{-4})\,.
\eeq
These branching ratios can be defined in terms of four independent parameters which contain all the available information:
\beq
\zeta_i &=& \frac{|V_{L/R}^{4i}|^2}{ \sum_{j=1}^3 |V_{L/R}^{4j}|^2}\,, \quad \sum_{i=1}^3 \zeta_i = 1\,, \label{eq:zeta}\\
\xi_V &=& \frac{\kappa_V^2 \Gamma_{V}^0}{ \sum_{V^{\prime }=W,Z,H} \kappa_{V^{\prime }}^2 \Gamma_{V^{\prime}}^0}\,, \quad \sum_{V=W,Z,H} \xi_V = 1\,; \label{eq:xi}
\eeq
so that
\beq
BR(T \to V q_i) = \zeta_i \xi_V\,.
\eeq

For experimental purposes, the decays into first or second generation cannot be distinguished: one can therefore express all the results 
in terms of the decay rates into light generations via $\zeta_{jet} = \zeta_1 + \zeta_2 = 1-\zeta_3$:
\beq
BR (T \to Z j) = \zeta_{jet} \xi_Z\,, & & BR (T \to Z t) = (1-\zeta_{jet}) \xi_Z\,, \\
BR(T \to H j) = \zeta_{jet} (1-\xi_Z-\xi_W)\,, & &  BR(T \to H t) = (1-\zeta_{jet})(1-\xi_Z-\xi_W)\,, \\
BR(T \to W^+ j) = \zeta_{jet} \xi_W\,, & & BR(T \to W^+ b) = (1-\zeta_{jet}) \xi_W\,.
\eeq
When studying pair production of $T$, which is dominated by model-independent QCD processes only sensitive to the mass of the VL quark, the phenomenology of the $T$ can be therefore completely described in terms of 4 independent parameters: the mass $M$, $\xi_W$, $\xi_Z$ and $\zeta_{jet}$.
As it will be clear later, single production processes may be sensitive to the separate values of $\zeta_1$ and $\zeta_2$, so the number of relevant parameters can be increased by one unit.

We can finally re-express the Lagrangian in Eq.(\ref{eq:topL}) in terms of the relevant 5 parameters as follows:
\begin{eqnarray}
\mathcal{L} &=&  \kappa_T \left\{ \sqrt{\frac{\zeta_i \xi_W}{\Gamma_W^0}} \frac{g}{\sqrt{2}}\; [\bar{T}_{L/R} W_\mu^+ \gamma^\mu d^i_{L/R} ]  +  \sqrt{\frac{\zeta_i \xi_Z}{\Gamma_Z^0}} \frac{g}{2 c_W} \; [\bar{T}_{L/R} Z_\mu \gamma^\mu u^i_{L/R} ]  \right.
\nonumber \\
&-& \left.  \sqrt{\frac{\zeta_i (1-\xi_Z-\xi_W)}{\Gamma_H^0}} \frac{M}{v}\; [\bar{T}_{R/L} H~u^i_{L/R} ] \right\} + h.c. \quad \mbox{with} \quad \zeta_3 = 1-\zeta_1 - \zeta_2\,. \label{eq:topL2}
\end{eqnarray}
The new parameter $\kappa_T$ is an overall coupling strength measure: it is not relevant for the branching ratios, nor for pair 
production (which is to a very good approximation due to QCD processes), however it will determine the strength of single production.
It can be written in terms of the parameters in the starting Lagrangian as
\beq
\kappa_T = \sqrt{\sum_{i=1}^3 |V_{L/R}^{4i} |^2} \sqrt{\sum_V \kappa_V^2 \Gamma_V^0}\,.
\eeq
It is important to notice that the $V_{L/R}^{4i}$ matrix elements are, in general, complex quantities as phases may be present in the mixing with light quarks. Since the parameters $\zeta_i$ are proportional to the square of mixing matrix entries, the information about phases is lost in the parametrisation in Eq.~(\ref{eq:topL2}). Such phases are crucial when considering, for instance, flavour bounds on couplings, however they will play a minor role in the LHC phenomenology which is the main focus of this parametrisation.
Phases are potentially relevant only in single production processes where interference terms give a sizeable contribution, which is not the case in the production modes we will consider in this work, as it will be clear in the following sections.

So far, we have completely and consistently neglected the contribution of the top mass both in the kinematic functions and in the suppressed couplings.
However, for VL quark masses below a TeV, the effects may be numerically relevant.
In Appendix~\ref{app:topmass}, we present a detailed discussion of the effects in all the relevant decay channels: one key point here is that the effect of the suppressed coupling, which is often dominant, introduces model dependence, therefore we would be forced to introduce new parameters in our Lagrangian.
On the other hand, we checked that in simple models such effects are always small, being below $10\div 20\%$ for $M = 600$ GeV (which is the level of present exclusion from direct searches at the LHC), therefore we will neglect their effect for now.
The only exception is the channel $T \to H t$: in this case, however, the sub-leading term is independent on the representation $T$ belongs to.
The latter coupling originates from the mass mixing between the VL quark and the SM top. Allowing for this mixing automatically generates such a coupling: a proof of this statement can be found in Appendix~\ref{app:topmass}.
It also turn out that the effect of the phase space is sub-dominant, and the main contribution comes from the new coupling.
For this reason, we suggest to complement the Lagrangian in Eq.~(\ref{eq:topL2}) with an additional term:
\beq
\Delta \mathcal{L}_{T\; single} =  - \kappa_T \sqrt{\frac{\zeta_3 (1-\xi_Z-\xi_W)}{\Gamma_H^0}} \frac{m_t}{v}\; [\bar{T}_{L/R} H~u^i_{R/L} ] + h.c.\,,
\eeq
where the new term has opposite chiralities compared to the one in Eq.~(\ref{eq:topL2}), and is suppressed by a factor $m_t/M$.
No extra free parameter needs to be introduced.
The addition of this term modifies the relation between the parameters $\zeta_i$ and $\xi_V$ with the branching ratios of the $T$:
\beq
BR (T \to H t) = \frac{\zeta_{3} \xi_H (1+\delta_H)}{1+ \zeta_3 \xi_H \delta_H}\,, 
\eeq
while for all other channels
\beq
BR (T \to V q_i) = \frac{\zeta_{i} \xi_V}{1+ \zeta_3 \xi_H \delta_H}\,.
\eeq
The correction $\delta_H$ is a simple function of the mass of the VL quark, and it is given by
\beq
\delta_H =   \frac{\lambda^{1/2} (1,\frac{m^2_H}{M^2}, \frac{m_t^2}{M^2})}{\left( 1-\frac{m^2_H}{M^2}\right)^2}  \left[ \left(1+\frac{m_t^2}{M^2}-\frac{m^2_H}{M^2} \right)\left(1+\frac{m^2_t}{M^2} \right) + 4 \frac{m^2_t}{M^2} \right] - 1 \sim 5 \frac{m^2_t}{M^2}\,, \label{eq:deltaH}
\eeq
where we expanded the result at leading order in $1/M^2$.
Numerically, this effect is $\delta_H \sim 39\%$ for $M = 600$ GeV, and it therefore leads to a substantial enhancement of the decay rate in $H t$.

\section{Complete and model independent parametrisation}
\label{sec:param}

In order to couple to the SM quarks, the new top partners must have the same colour as the standard ones 
(thus belonging to the fundamental representation of $SU(3)_c$) and have four possible charge assignments:
\beq
Q = 5/3 & \Rightarrow & X \to W^+ u^i\,; \nonumber \\
Q = 2/3 &  \Rightarrow & T \to W^+ d^i\,,\; Z u^i\,,\; H u^i\,; \nonumber\\ 
Q = -1/3 &  \Rightarrow & B \to W^- u^i\,,\; Z d^i\,,\; H d^i\,; \nonumber \\ 
Q = -4/3 &  \Rightarrow & Y \to W^- d^i\,; 
\label{eq:charges}
\eeq
where $i=1,2,3$ labels the 3 standard generations.
In order to simplify the analysis, we will limit ourselves to some reasonable and general assumptions on the couplings of the new 
quarks, mainly based on the search strategy followed by the LHC experiments.
In fact, the bounds on new states are always based on the assumption that only one new state contributes to the signal region.
This assumption can be easily satisfied in the case where, for each kind of new quark, there is a single mass eigenstate that is 
significantly lighter that the others. This statement can be specified in two distinct situations: either the mass of the second state is 
such that the production cross section can be safely neglected and thus it brings only a minor impact on the bound, or the mass splitting is such that 
the mixing between the two states is negligible. In the latter case, one can consider the contribution of the two (or more) states 
independently, and simply add the number of events generated by each resonance in the search bin.
This simple assumption can lead to significant simplification in the parametrisation, as discussed in the following section.

\subsection{Theoretical assumptions}
\label{subsec:param}

Following the search strategies, the main assumption we base our analysis on is that the signals given by the new states can be 
studied as independent, either due to a large mass splitting or negligible mixing effects.
The large mixing case can also be included under certain circumstances that will become clear at the end of this section.
The main assumptions leading to the effective Lagrangian we propose are summarised as follows:

\begin{itemize}
\item[-] the new quarks belong to {\it complete $SU(2)_{L}$ representations}: this assumption is justified by the necessity to have a model 
which is compatible with the gauge structure of the Standard Model. As a consequence, the mass splitting between different 
components of the multiplet, and also the mixing with the standard quarks as well as to heavier new quarks, are linked to the electroweak 
symmetry breaking sector.
\item[-] we assume a {\it Standard Model Higgs field}: therefore the mixing can only be generated by Yukawa-type interactions involving 
a doublet of $SU(2)_{L}$. This fact limits the size of the mass splitting which is related to the Higgs VEV, and the choice of quark representation~\cite{Cacciapaglia:2011fx}. 
Only 2 singlets, 3 doublets and 2 triplets are allowed, with varied hypercharge assignments.
This approximation is also valid for models with extended Higgs sector: in fact, while additional doublets do not change the structure of the mixings, other representations are forced to have a small vacuum expectation value to avoid too large corrections to the $\rho$ parameter, thus we can generically safely neglect their contribution to the mixing.
\item[-] the coupling to the standard quarks can involve either {\it left-} or {\it right-handed} quarks: once the representation the new states 
belong to is chosen, singlets and triplets can mix with the standard left-handed doublets, while the new doublets can only mix with the 
standard right-handed singlets.
\item[-] the chirality of the Yukawa couplings implies that the couplings to $W$, $Z$ and $H$ are also (predominantly) {\it chiral}. This is due to 
the fact that for singlets and triplets the mixing angles in the right-handed sector are suppressed with respect to the left-handed ones by 
the mass of the standard quarks over the new state mass~\cite{Cacciapaglia:2011fx}, while for doublets it's the left-handed mixings which are suppressed.
This stays true in models with more than one VL multiplet, as shown in Appendix~\ref{app:multiVL}, as long as the Yukawa couplings between VL multiplets are not too large.
\item[-] only decays to Standard Model quarks are allowed via standard gauge bosons. This assumption is justified by the small splitting between the masses of the components of the $SU(2)_{L}$ multiplet.
\end{itemize}

Complete $SU(2)_{L}$ representations, with the exception of singlets, will contain more than one VL quark with different charge: they decay into different final states, therefore the assumption of an isolated new quark is still viable.
If the final states are experimentally indistinguishable, one can always sum the two contributions to the same signal region.
Furthermore, decays of a VL quark into another, like for instance $X \to W^+ T$, are generically not allowed kinematically because the mass splitting between two states in the same multiplet, which is generated by the Higgs VEV, is typically much smaller than the $W$ mass.

\subsection{The effective Lagrangian}

The discussion in the previous section for the $T$ top partner, can be generalised to the other 3 kinds of VL quarks we are interested 
in this work. Therefore, the most complete effective model apt to describe their phenomenology would contain the following 4 sets of 
interactions:
\begin{multline}
\mathcal{L} =  \kappa_T \left\{ \sqrt{\frac{\zeta_i \xi_W^T}{\Gamma_W^0}} \frac{g}{\sqrt{2}}\; [\bar{T}_{L} W_\mu^+ \gamma^\mu d^i_{L} ]  +  \sqrt{\frac{\zeta_i \xi_Z^T}{\Gamma_Z^0}} \frac{g}{2 c_W} \; [\bar{T}_{L} Z_\mu \gamma^\mu u^i_{L} ]  \right.\\
\left. -  \sqrt{\frac{\zeta_i \xi_H^T}{\Gamma_H^0}} \frac{M}{v}\; [\bar{T}_{R} H u^i_{L}]  -  \sqrt{\frac{\zeta_3 \xi_H^T}{\Gamma_H^0}} \frac{m_t}{v}\; [\bar{T}_{L} H t_{R}] \right\}\\
 + \kappa_B \left\{ \sqrt{\frac{\zeta_i \xi_W^B}{\Gamma_W^0}} \frac{g}{\sqrt{2}}\; [\bar{B}_{L} W_\mu^- \gamma^\mu u^i_{L} ]  +  \sqrt{\frac{\zeta_i \xi_Z^B}{\Gamma_Z^0}} \frac{g}{2 c_W} \; [\bar{B}_{L} Z_\mu \gamma^\mu d^i_{L} ] -  \sqrt{\frac{\zeta_i \xi_H^B}{\Gamma_H^0}} \frac{M}{v}\; [\bar{B}_{R} H d^i_{L} ] \right\} \\
 + \kappa_X \left\{ \sqrt{\frac{\zeta_i}{\Gamma_W^0}} \frac{g}{\sqrt{2}}\; [\bar{X}_{L} W_\mu^+ \gamma^\mu u^i_{L} ]  \right\} + \kappa_Y \left\{ \sqrt{\frac{\zeta_i}{\Gamma_W^0}} \frac{g}{\sqrt{2}}\; [\bar{Y}_{L} W_\mu^- \gamma^\mu d^i_{L} ]  \right\} + h.c.\,, \label{eq:param}
\end{multline}
for leading left-handed mixing, while it suffices to exchange the chiralities L $\leftrightarrow$ R for leading right-handed coupling.
Note that $\xi_V^T$ and $\xi_V^B$ are in general different, also in models where the two VL quarks belong to the same representation.
In principle, the rates in the 3 generations may also be different, however this is not the case in the simplest cases.
As mentioned before, in typical models only one of the two mixings is large, and the other suppressed. This effective Lagrangian has 
been implemented in FeynRules~\cite{Christensen:2008py} for our analysis, and is described in more detail in 
Appendix~\ref{app:feynrules}.
The complete FeynRules files, together with the CalcHEP and MadGraph outputs, are available on the FeynRules website for the 
general model \cite{wwwFeynRules} and also for specific cases of a $T$ singlet, a SM-like doublet and a doublet with a $T$ and an 
exotic VL quark $X$ of charge 5/3  \cite{wwwFeynRules2}.
See also the website of the HEP model database project~\cite{hepmdb}.

The Lagrangian in Eq.(\ref{eq:param}) allows to express the decay rates in a simple and intuitive form:
\beq
BR(T \to W^+ j) = \frac{\zeta_{jet} \xi^T_W}{1+\zeta_3 \xi_H \delta_H}\,, & & BR(T \to W^+ b) = \frac{(1-\zeta_{jet}) \xi^T_W}{1+\zeta_3 
\xi_H \delta_H}\,,  \nonumber\\
BR (T \to Z j) = \frac{\zeta_{jet} \xi^T_Z}{1+\zeta_3 \xi_H \delta_H}\,, & & BR (T \to Z t) = \frac{(1-\zeta_{jet}) \xi^T_Z}{1+\zeta_3 
\xi_H \delta_H}\,,\\
BR(T \to H j) = \frac{\zeta_{jet} (1-\xi^T_Z-\xi^T_W)}{1+\zeta_3 \xi_H \delta_H}\,, & &  BR(T \to H t) = \frac{(1-\zeta_{jet})
(1-\xi^T_Z-\xi^T_W) (1+\delta_H)}{1+\zeta_3 \xi_H \delta_H}\,, \nonumber\\
 & & \nonumber \\
BR(B \to W^- j) = \zeta_{jet} \xi^B_W\,, & & BR(B \to W^- t) = (1-\zeta_{jet}) \xi^B_W\,, \nonumber \\
BR (B \to Z j) = \zeta_{jet} \xi^B_Z\,, & & BR (B \to Z b) = (1-\zeta_{jet}) \xi^B_Z\,,  \\
BR(B \to H j) = \zeta_{jet} (1-\xi^B_Z-\xi^B_W)\,, & &  BR(B \to H b) = (1-\zeta_{jet})(1-\xi^B_Z-\xi^B_W)\,,\nonumber
\eeq
so that the BR of the top and bottom partner only depend on 3 parameters each ($\zeta_{jet}$, $\xi_W^{B/T}$ and $\xi_Z^{B/T}$), 
while $\delta_H$ is a known function of $M$ given in Eq.~(\ref{eq:deltaH}).
For the exotic-charge VL quarks:
\beq
BR(X \to W^+ j) = \zeta_{jet}\,, & & BR(X \to W^+ t) = (1-\zeta_{jet})\,, \\
 & & \nonumber \\
BR(Y \to W^- j) = \zeta_{jet}\,, & & BR(Y \to W^- b) = (1-\zeta_{jet})\,,
\eeq
so that they depend on a single parameter each, $\zeta_{jet}$.
As we already discussed, in the formulas above we neglected the top mass, except for the channel $T \to H t$ where large model-independent corrections are expected.  
In the other potentially affected channel, i.e. $T \to W b$, $Z t$, $B \to W t$ and $X \to W t$, this approximation is numerically sensible for the range of masses LHC will be probing, and more details on the model-dependent top mass corrections can be found in Appendix~\ref{app:topmass}.
The above formulas for the BR will mainly be used to determine the parameters we propose, $\zeta_i$ and $\xi_V$, starting from the physical branching ratios in a specific model containing VL quarks.

The mass of the VL quark will determine its production rates, especially for pair production which is dominated by QCD processes. The coupling strength factors $\kappa_Q$ will drive the electroweak pair and single production cross sections, which are therefore sensitive to the overall strength of the coupling, similarly to the single top production processes in the Standard Model.

\subsection{Benchmark scenarios from flavour bounds}
\label{sect:benchmarks}

As these parameters will play a crucial role in the phenomenology, it is important to have a handle on the reasonable value they may have.
The mixing between the VL quarks and the standard ones is generated by Yukawa-type interactions, therefore such mixing will also modify the diagonalisation of the standard Yukawa matrices and potentially generate tree level flavour-changing couplings to the Z boson.
The reason for this is the vector-like nature of the new quarks: either the left-handed or the right-handed chirality (or both) of the new states will differ from the Standard Model quark ones. Therefore the presence of $Z$-mediated FCNC's is inevitable, unless tuned cancellations occur.
Such cancellations may take place naturally in models where flavour~\cite{Redi:2011zi} and/or custodial symmetries~\cite{Agashe:2006at} are built-in: however, such models necessarily contain more than one VL quark, and their phenomenology may be more complex that the one presented here.
In such cases, dedicated model-dependent searches may be necessary at the LHC.

In all extensions of the Standard Model, the only consistent way to add VL quarks, independently on their number, is to introduce new 
complete $SU(2)_{L}$ representations. In the gauge basis, therefore, gauge interactions are diagonal and completely determined by 
the quantum numbers of the representations. After the diagonalisation of the mass, mixing matrices will induce off-diagonal couplings 
in the form of Eq.(\ref{eq:topL}) - this is a simple generalisation of the origin of the flavour-changing $W$ couplings in the Standard 
Model. Thus, the physics of the new states can be completely encoded in mixing matrices $V_L$ and $V_R$ describing the mixing 
between standard and new quarks.
The simplest scenario contains a single VL representation~\cite{delAguila:2000rc,Cacciapaglia:2010vn}, and in this section we will use this as a toy 
model to study the flavour properties of a more general scenario. The case of a new singlet was previously discussed 
in \cite{AguilarSaavedra:2002kr}. The 7 possibilities (2 singlets, 3 doublets and 2 triplets) can be recast in terms of our 
parametrisation. The branching ratios into the 3 generations can be expressed in terms of the mixing matrices as in (\ref{eq:zeta}), 
reported here for clarity:
\beq
\zeta_i = \frac{|V_{L/R}^{4i}|^2}{ \sum_{j=1}^3 |V_{L/R}^{4j}|^2}\,, \quad \sum_{i=1}^3 \zeta_i = 1\,;
\eeq
while the branching into bosons (\ref{eq:xi}) is determined by the representation.
The results are given in Tab.~\ref{tab:1VLpar}, where each representation is identified by $(n, Y)$, where $n$ is the dimension of the SU(2) $n$-plet (2 for a doublet, and so on) and $Y$ is the hypercharge. 
The simplicity of this parametrisation is based on the fact that only one set of Yukawa couplings generates all the new mixings.
For the doublet with standard hypercharge ($Y=1/6$) the situation is more complex, because there are two possible Yukawa couplings: one involving the up-singlets ($\lambda_u$) and one with the down-singlets ($\lambda_d$).
Therefore, in general, such simple re-parametrisation is not possible.
We can use our parametrisation only in 3 simple limits: when one of the two new Yukawas is set to zero or small, or when they are equal.
The parametrisations in the 3 cases are listed in Tab.~\ref{tab:1VLparD}.

\begin{table}[tb]
\begin{center}
\begin{tabular}{c|ccc|ccc|c|c}
\toprule
& \multicolumn{3}{c|}{T} & \multicolumn{3}{c|}{B} & \multicolumn{1}{c|}{X} & \multicolumn{1}{c}{Y} \\
& $\xi_W^T$ &  $\xi_Z^T$ & $\kappa_T/\kappa$ & $\xi_W^B$ &  $\xi_Z^B$ & $\kappa_B/\kappa$ & $\kappa_X/\kappa$ & $\kappa_Y/\kappa$ \\
\midrule
$(1, 2/3)$ & $\frac{1}{2} + \frac{1}{4} \epsilon_H$ & $\frac{1}{4} + \frac{1}{8} \epsilon_H$ & $\sqrt{2-\epsilon_H}$ & & & & & \\ 
$(1, -1/3)$ & & & & $\frac{1}{2} + \frac{1}{4} \epsilon_H$ & $\frac{1}{4} + \frac{1}{8} \epsilon_H$ & $\sqrt{2-\epsilon_H}$ & & \\ 
\midrule
$(2, 7/6)$ & $0$ & $\frac{1}{2} + \frac{1}{2} \epsilon_H$ & $\sqrt{1-\epsilon_H}$ & & & & $1$ & \\ 
$(2, -5/6)$ & & & & $0$ & $\frac{1}{2} + \frac{1}{2} \epsilon_H$ & $\sqrt{1-\epsilon_H}$ & & $1$ \\ 
\midrule
$(3, 2/3)$ & $\frac{1}{2} + \frac{1}{4} \epsilon_H$ & $\frac{1}{4} + \frac{1}{8} \epsilon_H$ & $\sqrt{2-\epsilon_H}$ & $0$ & $\frac{1}{2} + \frac{1}{2} \epsilon_H$ & $\sqrt{2-2 \epsilon_H}$ & $\sqrt{2}$ &  \\ 
$(3, -1/3)$ & $0$ & $\frac{1}{2} + \frac{1}{2} \epsilon_H$ & $\sqrt{1 - \epsilon_H}$ &$\frac{1}{2} + \frac{1}{4} \epsilon_H$ & $\frac{1}{4} + \frac{1}{8} \epsilon_H$ & $\sqrt{1-\frac{1}{2} \epsilon_H}$ &  & $1$ \\ 
\bottomrule
\end{tabular}
\caption{Branching ratios into gauge bosons in the case of a single VL representation, where $(n, Y)$ labels a SU(2) $n$-plet with hypercharge $Y$. Here we list the results at leading order in $1/M^2$, where $\Gamma_W^0 \sim 2 \Gamma_Z^0 \sim 1$, and $\Gamma_H^0 \sim 1/2 - \epsilon_H = 1/2 - \frac{m_H^2}{M^2}$. The coupling strengths are proportional to $\kappa = \sqrt{\sum_{j=1}^3 |V_{L/R}^{4j}|^2}$.} \label{tab:1VLpar}
\end{center}
\end{table}

The same mechanism that generates the off diagonal couplings responsible for the decays of the VL quarks will also generate flavour changing couplings for the $W$, $Z$ and Higgs.
The most dangerous ones are the ones involving the $Z$, because they are absent in the Standard Model at tree level, thus very strong bounds on such couplings derive from flavour observables. The couplings of the Higgs involve both right and left-handed mixing matrices and, as mentioned above, one of the two will be suppressed by the mass of the light quarks, thus in general the flavour changing Higgs couplings are sufficiently suppressed.
In the gauge basis, the $Z$ couplings for up and down quarks (considering in general $n$ VL partners $\psi_i$) can be written as~\cite{Cacciapaglia:2011fx}:
\beq
g_{Z\bar{q}q}^{gauge} = \frac{g}{c_W} \left( T_3^{sm} - Q s_W^2 \right) \delta^{ij} + \frac{g}{2 c_W} \left( \begin{array}{cccccc}
0 & & & & & \\
& 0 & & & & \\
& & 0 & & & \\
& & & 2 (T_3^{\psi_1} - T_3^{sm}) & & \\
& & & & \ddots & \\
& & & & & 2 (T_3^{\psi_n} - T_3^{sm})
\end{array} \right)\,.
\eeq
The first term, proportional to the identity matrix, is the coupling of the standard quark (up or down type, $T_3^{sm} = \pm 1/2$ or $0$).
From the equation it is clear that the non-standard couplings are due to the weak isospin of the VL quarks, which must be different from the standard ones at least for one chirality.
Once going to the mass eigenstate basis via the unitary matrices $V_{L/R}$, the first terms stay untouched, while the second term, sensitive to the difference in isospin, will generate off-diagonal couplings:
\beq
(g_{Z\bar{q}q}^{mass})_{ij} = \frac{g}{c_W} \left( T_3^{sm} - Q s_W^2 \right) \delta^{ij} + \frac{g}{2 c_W} \left( \sum_{k = 4}^n   2 (T_3^{\psi_k} - T_3^{sm})  V_{L/R}^{*,ki} V_{L/R}^{kj}  \right)\,.
\label{eq:Zcoups}
\eeq
The second term here is responsible for the generation of both the decays of the VL quarks and the off-diagonal $Z$ couplings between the standard quarks, and both couplings are proportional to the same matrix elements: $V_{L/R}^{ki}$ where $k$ spans over the VL quarks.

\begin{table}[tb]
\begin{center}
\begin{tabular}{l|ccc|ccc}
\toprule
$(2, 1/6)$ & \multicolumn{3}{c|}{T} & \multicolumn{3}{c}{B}  \\
 & $\xi_W^T$ &  $\xi_Z^T$ & $\kappa_T/\kappa$ & $\xi_W^B$ &  $\xi_Z^B$ & $\kappa_B/\kappa$ \\
\midrule
$\lambda_d=0$   & $0$ & $\frac{1}{2} + \frac{1}{2} \epsilon_H$ & $\sqrt{1-\epsilon_H}$ & $1$ & $0$ & $1$ \\ 
 \midrule
$\lambda_u = 0$   & $1$ & $0$ & $1$ & $0$ & $\frac{1}{2} + \frac{1}{2} \epsilon_H$ & $\sqrt{1-\epsilon_H}$ \\ 
 \midrule
 $\lambda_u = \lambda_d$   & $\frac{1}{2} + \frac{1}{4} \epsilon_H$ & $\frac{1}{4} + \frac{1}{8} \epsilon_H$ & $\sqrt{2-\epsilon_H}$& $\frac{1}{2} + \frac{1}{4} \epsilon_H$ & $\frac{1}{4} + \frac{1}{8} \epsilon_H$ & $\sqrt{2-\epsilon_H}$ \\ 
\bottomrule
\end{tabular}
\caption{Branching ratios into gauge bosons for a doublet with standard hypercharge, in three limits: when one of the two Yukawa couplings is zero, or when they are equal.} \label{tab:1VLparD}
\end{center}
\end{table}

In the case of a single VL representation, one can easily be convinced that the couplings with the largest mixing matrices (the other being suppressed by the mass of the standard quarks) has $2 (T_3^{\psi_k} - T_3^{sm}) = \pm 1$.
For instance, for singlets ($T^\psi_3 = 0$), the large mixing takes place in the left-handed sector, where the standard quarks have isospin $\pm 1/2$.
Using the re-parametrisation in terms of branching ratios, the couplings of the $Z$ to the standard quarks can be rewritten as (ignoring phases):
\beq
\left(g_{Z\bar{q}q}^{sm} \right)_{ij}= \frac{g}{c_W} \left( T_3^{sm} - Q s_W^2 \right) \delta^{ij} \pm \frac{g}{2 c_W} \left( \kappa^2 \sqrt{\zeta_i \zeta_j}  \right)\,, \quad i, j = 1,2,3\,.
\eeq
The off-diagonal terms can therefore pose a bound on the product of two $\kappa \sqrt{\zeta_i}$ terms.
As the $Z$ is much heavier that any mesons, it is useful to parametrise the bounds in terms of effective 4-fermion operators after integrating out the $Z$:
\beq
\frac{g^2 \kappa^4}{4 m_W^2} \sqrt{\zeta_{i1} \zeta_{i2} \zeta_{j1} \zeta_{j2}} \; (\bar{q}_{i1} \gamma_\mu q_{i2} )\, (\bar{q'}_{j1} \gamma_\mu q'_{j2} )\,,
\eeq
where $q$ and $q'$ are up or down type quarks of the same chirality.
To estimate the bounds from flavour, we can compare the expressions for our operators with the bounds in Ref.~\cite{Isidori:2010kg}, which involve $\Delta F = 2$ operators in the left-handed sector (similar bounds apply to the right-handed operators):
\beq
(\bar{s}_L \gamma^\mu d_L)^2 & \Rightarrow |c| < 9.0 \cdot 10^{-7} & \Rightarrow \kappa^4 \zeta_1 \zeta_2 < 5.5 \cdot 10^{-8} \quad (\kappa < 0.015/(\zeta_1\zeta_2)^{1/4})\,, \\
(\bar{b}_L \gamma^\mu d_L)^2 & \Rightarrow |c| < 3.3 \cdot 10^{-6} & \Rightarrow \kappa^4 \zeta_1 \zeta_3 < 2.0 \cdot 10^{-7} \quad (\kappa < 0.02/(\zeta_1\zeta_3)^{1/4})\,, \\
(\bar{b}_L \gamma^\mu s_L)^2 & \Rightarrow |c| < 7.6 \cdot 10^{-5} & \Rightarrow \kappa^4 \zeta_1 \zeta_2 < 4.6 \cdot 10^{-6} \quad (\kappa < 0.045/(\zeta_2\zeta_3)^{1/4})\,, \\
(\bar{c}_L \gamma^\mu u_L)^2 & \Rightarrow |c| < 5.6 \cdot 10^{-7} & \Rightarrow \kappa^4 \zeta_1 \zeta_2 < 3.4 \cdot 10^{-8} \quad (\kappa < 0.014/(\zeta_1\zeta_2)^{1/4})\,.
\eeq
Here, the bounds on the coefficient $|c|$ assume a scale of 1 TeV, furthermore we only considered the bounds on the real parts as bounds on the imaginary parts of the coefficient are much stronger and would mainly affect the phases of the mixing matrices (that we ignored in our parametrisation).
From such expressions we can learn two things: the flavour bounds only apply to the product of the coupling to two generations, therefore the bounds can be evaded by coupling the VL quarks mainly to one generation. 
Furthermore, in the down-sector, there are strong bounds on all three combinations, therefore we can conclude that in the presence of a VL down partner ($B$ and $Y$), the coupling with a single generation is preferred.
For the up-sector, there are no bounds involving the third generation, therefore one can evade bounds by allowing sizeable couplings with the top and one of the two light generations: this case applies for representations that do not contain a $B$ partners (i.e., only $T$ and $X$).
The coupling strength $\kappa$ does not receive constraining bounds in this case: strong bounds are obtained if sizeable mixing with two generations are attained.
In the case of maximal mixing (i.e. $\zeta_i = \zeta_j = 1/2$), the bounds range $\kappa < 0.02 \div 0.06$.

The mixing with the VL quarks also affects the diagonal couplings of the $Z$ to the standard quarks: in this case, the correction is of the order
\beq
\delta g_{Z\bar{q}_i q_i} = \pm \frac{g}{2 c_W} \kappa^2 \zeta_i\,.
\eeq
The couplings of the $Z$ to light quarks have been precisely measured at LEP~\cite{ALEPH:2005ab} and other low energy experiments, and they are proportional to the branching ratio to the given generation.
This implies that these bounds allow to extract an absolute bound on the coupling strength $\kappa$.
This story is true for all quarks, except for the top whose couplings to the $Z$ are not known.
In this case, therefore, the only bound comes from loop corrections (mainly the $T$ parameter) and corrections to the $W$ coupling to the bottom, that mediates its decays.
To extract bounds on the parameters, we take a very conservative approach: we assume that only one coupling is affected and compare the correction $\delta g_{Z\bar{q}_i q_i}$ to the error on the measurement, assuming therefore that the central value of the measurement agrees with the Standard Model prediction.
A more detailed analysis would require a new complete fit of the electroweak precision measurements, which is beyond the scope of this paper and sensitive to the details of the model.
The bounds on $\kappa$ from modifications of the $Z$ couplings are listed below:
\beq
Z \bar{u} u\, (APV) & \Rightarrow & |\delta g_{L/R}| < 3\times 0.00069 \rightarrow \kappa < 0.074/\sqrt{\zeta_1}\,,   \\
Z \bar{d} d\, (APV) & \Rightarrow & |\delta g_{L/R}| < 3\times 0.00062 \rightarrow \kappa < 0.07/\sqrt{\zeta_1}\,, \\
Z \bar{s} s\, (LEP)  & \Rightarrow & \begin{array}{c}  |\delta g_L| < 3 \times 0.012 \rightarrow \kappa < 0.3/\sqrt{\zeta_2}\,, \\  |\delta g_R| < 3\times 0.05 \rightarrow \kappa < 0.6/\sqrt{\zeta_2}\,, \end{array} \\
Z \bar{c} c\, (LEP) &  \Rightarrow & \begin{array}{c}  |\delta g_L| < 3 \times 0.0036 \rightarrow \kappa < 0.17/\sqrt{\zeta_2}\,, \\  |\delta g_R| < 3\times 0.0051 \rightarrow \kappa < 0.20/\sqrt{\zeta_2}\,, \end{array} \\
Z \bar{b} b\, (LEP) &  \Rightarrow & \begin{array}{c} |\delta g_L| < 3\times 0.0015 \rightarrow \kappa < 0.11/\sqrt{\zeta_3}\,, \\  |\delta g_R| < 3\times 0.0063 \rightarrow \kappa < 0.23/\sqrt{\zeta_3}\,, \end{array} \\
Z \bar{t} t\, (T,\; \delta g_{Wtb}) &  \Rightarrow &  \kappa < 0.1\div0.3/\sqrt{\zeta_3}\,.
\label{eq:kWtb}
\eeq
The best bounds on the first generation couplings are coming from the measurement of the weak charge of the Cesium atom (Atomic 
Parity Violation)~\cite{Deandrea:1997wk}, while the others are determined from the LEP measurements of the hadronic cross sections 
and asymmetries. The bound in formula \ref{eq:kWtb} is obtained from the electroweak precision tests as in \cite{Cacciapaglia:2010vn}.
The flavour diagonal bounds on $\kappa$ are about one order of magnitude milder than the flavour violating ones.

From this simple analysis, we can derive a set of benchmark models than can be used to reduce the number of parameters in the first studies:
\begin{itemize}
\item[-] in the presence of a bottom partner ($B$ and $Y$), only the coupling to one family is allowed to be large: we therefore have 3 benchmark models with $\zeta_1 = 1$ ($\kappa \lesssim 0.07$), $\zeta_2 = 1$ ($\kappa \lesssim 0.2$) and $\zeta_3 = 1$;
\item[-] in the absence of a bottom partner (thus only $T$ and $X$), one can allow for couplings to two generations: $\zeta_2 = 0$ ($\kappa \lesssim 0.1$) and $\zeta_1 = 0$ ($\kappa \lesssim 0.3$).
\end{itemize}
A scenario with significant couplings to all generations is also allowed, however with an extra order of magnitude suppression on $\kappa$: in this case, single production will yield very small cross sections, and its relevance postponed to higher mass values.
Note also that the bounds on the couplings have been extracted in a specific scenario (a single light VL representation), while the bounds may be weakened in more involved models, therefore they should only be considered as guiding points without limiting the validity of the parametrisation to more general scenarios.

\section{Production processes}
\label{sec:prod}

The production cross sections of VL quarks can be grouped in five classes:
\begin{itemize}
\item[-] {\it pair production}: this class is largely dominated by QCD production, which is model independent as it only depends on the mass of the new fermion;
\item[-] {\it single production in association with tops};
\item[-] {\it single production in association with jets (where jet denotes any light quark)};
\item[-] {\it single production in association with a boson}: including $W^\pm$, $Z$ and the Higgs $H$;
\item[-] {\it mono production through gluon chromomagnetic coupling} via a loop-induced operator: such operator will also contribute to the production with jets (and tops).
\end{itemize}
Our proposal allows to write the production cross sections explicitly in terms of model-independent cross sections multiplied by the parameters that also enter the branching ratios, an approach similar to the one followed in~\cite{DeSimone:2012fs}.
In this way, one can study all models at once by computing the efficiencies of each search in various channels, and also, given a model, correlate observations in various channels.
In the following we will discuss the 5 production mechanisms separately.

\subsection{Pair production}

Pair production is dominated by QCD production via gluons:
\beq
\bar{q} q, g g \to \bar{Q} Q\,.
\eeq
The cross section is model independent as it only depends on the mass of the VL quark, and it decreases quickly for higher masses due to PDF suppression, as shown in Tab.~\ref{tab:pairprod}.
There are also contributions from electroweak gauge bosons, which are sub-leading in terms of cross section.
Production via neutral currents ($Z$ and $\gamma$) have the same final states as QCD production, thus they are completely negligible.

\begin{table}[tb]
\begin{center}
\begin{tabular}{l|c|c|c}
\toprule
 & $\bar{Q} Q$ (QCD) & $\bar{Q} Q'\; (W^+)$ &  $\bar{Q} Q'\; (W^-)$  \\
\midrule
$M = 600$ GeV & $109 (167)$ & $3.95$ & $1.12$ \\
$M = 800$ GeV & $14.3 (20.5)$ & $0.646$ & $0.165$ \\
$M = 1000$ GeV & $2.37 (3.24)$ & $0.119$ & $0.0285$ \\
\bottomrule
\end{tabular}
\caption{Pair production cross sections (in fb) at 8 TeV for processes dominated by QCD, and for s-channel $W$ exchange ($T_{ij}^\pm= 1$, i.e. for a doublet). The values have been computed at LO with MadGraph, while the values in brackets are NLO+NNLL results from~\cite{Cacciari:2011hy} with the MSTW2008 PDF set} \label{tab:pairprod}
\end{center}
\end{table}

Production via a $W$ boson, on the other hand, can give rise to potentially interesting channels like:
\beq
\bar{q} q' &\to& W^+ \to \bar{T} X, \bar{B} T, \bar{Y} B \label{eq:offdiagonalplus}\\
\bar{q} q' &\to& W^- \to \bar{X} T, \bar{T} B, \bar{B} Y \label{eq:offdiagonalminus}
\eeq
Such cross sections are however model-dependent, as they depend on the
representation the VL quarks belong to. For definiteness, one can
parametrise the coupling such that: 
\begin{equation}
\mathcal{L}=\frac{g}{\sqrt{2}}\;[\bar{Q}^{i}W_{\mu }^{\pm }\gamma ^{\mu
}T_{ij}^{+}Q^{j\prime }]+h.c.\,;\label{eq:offdiagonallagrangian}
\end{equation}%
where $T_{ij}^{+}$ depends on the representation of $SU(2)_{L}$ the two VL
quarks belong to. For a $N$-plet: 
\begin{equation}
T_{ij}^{+}=%
\begin{bmatrix}
0 & c_{1} & 0 & \ldots  & 0 \\ 
0 & 0 & c_{2} & \ldots  & 0 \\ 
\vdots  & \vdots  & \vdots  & \ddots  & \vdots  \\ 
0 & 0 & 0 & \ldots  & c_{N-1} \\ 
0 & 0 & 0 & \ldots  & 0%
\end{bmatrix}%
,\text{ \ }T_{ij}^{-}=(T_{ij}^{+})^{\dag },\text{ \ \ }c_{k}=\sqrt{%
k(N-1)-k(k-1)} \label{eq:offdiagonalcoupling}
\end{equation}%
where $k=1,...,N-1$ \cite{GriffithsQM}. Therefore, the $W$-mediated pair production can be
parametrised as 
\begin{equation}
\sigma _{W}(\bar{Q}^{i}Q^{j\prime })=T_{ij}^{\pm 2}\text{ }\bar{\sigma}%
_{W}^{\pm }\,,\label{eq:offdiagonalsigma}
\end{equation}%
where the sign refers to the sign of the $W$ in the s-channel. The numerical values for the cross sections of the processes in Eqs. (\ref{eq:offdiagonalplus}) and (\ref{eq:offdiagonalminus}) are listed in Tab.~\ref{tab:pairprod}. 
The electroweak cross sections are very small and also strongly suppressed for large masses, therefore their impact on the search strategies can be safely neglected.\\

Another potentially relevant production process is represented by the production of a pair of VL quarks $Q Q'$, mediated by a $W$, $Z$ or Higgs in the $t$-channel. This process is completely absent in QCD and, depending on subsequent decays, it can give rise to final states with peculiar kinematics or same-sign dileptons. It must be stressed however that this channel is proportional to $\kappa_Q^4$, because it requires both couplings of the gauge boson to be from Eq.~(\ref{eq:param}), therefore rates are expected to be fairly small in realistic scenarios due to flavour bounds. 
Nevertheless, we will discuss these processes because of their peculiar final states, and because they can give rise to large cross sections in models where large mixings with the first generation are allowed, like models with more than one VL multiplet where the corrections to the $Z$ couplings can be partially cancelled.
This kind of process can give rise to the following final states:
\beq
TT\,, \quad BB\,, \quad XB\,, \quad TB\,, \quad TY\,.
\eeq
The same-sign final states, $TT$ and $BB$, can be mediated by a t-channel exchange of the two neutral bosons, and are present in all models because they involve a single VL quark.
Their cross sections can be expanded as in terms where we factor out powers of the parameters of our Lagrangian, the $\zeta_i$ and $\xi_V$, times coefficients that only depend on the mass of the VL quarks being produced:
\beq
\sigma (TT) & = & \kappa_T^4 \left( (\xi_Z^T)^2 \sum_{i,j=1}^2 \zeta_i \zeta_j \bar{\sigma}_{Zij}^{TT} + (\xi_H^T)^2 \sum_{i,j=1}^2 \zeta_i \zeta_j \bar{\sigma}_{Hij}^{TT} \right)\,; \\
\sigma (BB) & = & \kappa_B^4 \left( (\xi_Z^B)^2 \sum_{i,j=1}^3 \zeta_i \zeta_j \bar{\sigma}_{Zij}^{BB} + (\xi_H^B)^2 \sum_{i,j=1}^3 \zeta_i \zeta_j \bar{\sigma}_{Hij}^{BB} \right)\,.
\eeq
Such cross sections are symmetric in the initial states ($u_i u_i$ for $TT$, and $d_i d_j$ for $BB$), so that $\bar{\sigma}_{12} = \bar{\sigma}_{21}$.
The remaining 3 processes are only present if both VL quarks are present in the theory, and if both couple to SM quarks: $XB$ and $TY$ are both mediated by a $W$ exchange and are initiated by a $u_i u_j$ and $d_i d_j$ initial state respectively, while $TB$ is mediated by all 3 bosons and initiated by $u_i d_j$. 
These cross sections can also be expanded in terms of model-independent coefficients: in this expansion for $QQ'$, the first flavour index $i$ refers to the coupling of the first VL quark $Q$, the second $j$ to $Q'$, so the coefficients are not symmetric in flavours:
\beq
\sigma (TB) & = & \kappa_T^2 \kappa_B^2  \left( \xi_Z^T \xi_Z^B \sum_{i=1}^2 \sum_{j=1}^3 \zeta_i \zeta_j \bar{\sigma}_{Zij}^{TB} +  \xi_H^T \xi_H^B \sum_{i=1}^2 \sum_{j=1}^3 \zeta_i \zeta_j \bar{\sigma}_{Hij}^{TB} +  \xi_W^T \xi_W^B \sum_{i=1}^3 \sum_{j=1}^2 \zeta_i \zeta_j \bar{\sigma}_{Wij}^{TB} \right)\,; \\
\sigma (XB) & = & \kappa_X^2 \kappa_B^2 \left( \xi_W^B \sum_{i=1}^2 \sum_{j=1}^2 \zeta_i \zeta_j \bar{\sigma}_{Wij}^{XB}  \right)\,, \\
\sigma (TY) & = & \kappa_T^2 \kappa_Y^2 \left( \xi_W^T \sum_{i=1}^3 \sum_{j=1}^3 \zeta_i \zeta_j \bar{\sigma}_{Wij}^{TY}  \right)\,.
\eeq
Some terms are missing in the sum because of the absence of the top in the proton, and this structure will be preserved once QCD corrections in $\alpha_s$ are included, so that the validity of this expansion goes beyond leading order.
Furthermore, we systematically neglected potential interference between different bosons in the $t$-channel: we numerically checked that the impact of such interference terms amounts to a few percent at most, so that they can be neglected at this point to simplify the expansions.
The largest contribution to the cross sections are due to the mixing with first generation, because it corresponds to processes initiated by two valence quarks: we found that this is indeed true numerically, so that the contribution of the other two generations (and of anti-quarks) can be ignored for the phenomenology.
For illustration of the relevance of these processes, in Tab.~\ref{tab:EWsamesign}, we listed the coefficients for the first generation for a fixed masses $M = 600$, $800$ and $1000$ GeV and at a centre of mass energy of 8 TeV.
More complete tables can be found in the Appendix~\ref{app:feynrules}, including the effect of mixing with second and third generation.
We can anticipate that the cross sections will decrease with the mass of the VL quarks slower than QCD pair production, because of the valence-quark initiated processes, thus making them an attractive channel at the LHC.
Furthermore, the overall size of the cross sections scales like the coupling factor $\kappa_Q^4$, thus it it very sensitive to the size of the mixing.
For the benchmark model with mixing with the first generation only, this means that the coefficients in Tab.~\ref{tab:EWsamesign} must be weighted by a factor of $\kappa_Q^4 < (0.07)^4 = 2\cdot 10^{-5}$.
On the other hand, larger mixing can easily give rise to cross sections of several fb.

\begin{table}[tb]
\begin{center}
\begin{tabular}{c||cc||cc||ccc||c||c}
\toprule
& \multicolumn{2}{c||}{$TT$}  & \multicolumn{2}{c||}{$BB$}      & \multicolumn{3}{c||}{$TB$} & $XB$ & $TY$     \\
\midrule
& $\bar \sigma_{Z11}^{TT}$ & $\bar \sigma_{H11}^{TT}$ & $\bar \sigma_{Z11}^{BB}$ & $\bar \sigma_{H11}^{BB}$ & $\bar \sigma_{W11}^{TB}$ & $\bar \sigma_{Z11}^{TB}$ & $\bar \sigma_{H11}^{TB}$ & $\bar \sigma_{W11}^{XB}$ & $\bar \sigma_{W11}^{TY}$  \\
\midrule
$M = 600$ GeV & 86500 & 80600 & 16900 & 15700 & 70700 & 73300 & 67800 & 166900 & 32800 \\
$M = 800$ GeV & 66900 & 64100 & 10400 & 10100 & 50400 & 51000 & 49400 & 131300 & 20500 \\
$M = 1000$ GeV & 45900 & 44900 & 5700 &  5700 & 31300 & 31600 & 31100 &  89400 & 11300 \\
\bottomrule
\end{tabular}
\caption{Coefficients for $QQ^{(\prime)}$ pair production cross sections (in fb) at 8 TeV for different values of the VL masses, and listing only the coefficients of the first generation. The values have been computed at LO with MadGraph.} \label{tab:EWsamesign}
\end{center}
\end{table}

\subsection{Single production with tops and with jets}

These final states can be obtained, at leading order (LO), by the exchange in $t$ or $s$-channel of a $W$ and/or a $Z$ boson(s), due to the presence of the couplings $\xi_W$ and/or $\xi_Z$ in the Lagrangian in Eq.(\ref{eq:param}). Contributions of the Higgs will always be suppressed by the small masses of the light quarks.
We can therefore expand the total production cross sections by factoring out factors of $\zeta_i$ and $\xi_V$ (and an overall factor $\kappa_Q^2$), with coefficients that are model independent as they only depend on the mass of the VL quark via the kinematics.
Here we will neglect contributions that are suppressed by extra factors of $\kappa_Q$, thus we cut the expansion to the leading $\kappa_Q^2$ terms.
The cross sections for processes with a single top and a single VL quark in the final state can therefore be expanded as:
\beq
\sigma (T \bar{t} + \bar{T} t) &=& \kappa_T^2 \left( \xi_Z  \zeta_3\; (\bar{\sigma}^{T \bar{t}}_{Z3}+\bar{\sigma}^{\bar T t}_{Z3}) + \xi_W \sum_{i=1}^3 \zeta_i\; (\bar{\sigma}^{T\bar{t}}_{Wi} + \bar{\sigma}^{\bar T t}_{Wi}) 
\right)
\,, \\
\sigma (B t + \bar{B} \bar{t}) &=& \kappa_B^2 \left(\xi_W \sum_{i=1}^2 \zeta_i\; ( \bar{\sigma}^{B t}_{Wi} + \bar{\sigma}^{\bar B \bar t}_{Wi})\right)\,, \\
\sigma (B \bar{t} + \bar{B} t) &=& \kappa_B^2 \left( \xi_W  \zeta_3\; (\bar{\sigma}^{B \bar{t}}_{W3} + \bar{\sigma}^{\bar B t}_{W3}) \right)\,,\\
\sigma (X \bar{t} +\bar{X} t ) &=& \kappa_X^2 \left( \xi_W \sum_{i=1}^2 \zeta_i\; ( \bar{\sigma}^{X \bar{t}}_{i} + \bar{\sigma}^{\bar X t}_{i} ) \right)\,, \\
\sigma (Y t + \bar{Y} \bar{t}) &=& \kappa_Y^2 \left( \xi_W \sum_{i=1}^3 \zeta_i\; ( \bar{\sigma}^{Y t}_{i} + \bar{\sigma}^{\bar Y \bar t}_{i} )\right)\,.
\eeq
In these formulae we neglect the interference between $W$ and $Z$ exchange: in fact, we have verified numerically that they are small, the main reason being that interference is present only for a limited number of diagrams and between an $s$- and $t$-channel exchange.
Quantitatively, the interference terms are always below a percent level, therefore the approximation is very accurate and allows for a great simplification of the formulae.
This expansion is also valid once QCD corrections are included, and the effect can be completely included in the model-independent $\bar{\sigma}$ coefficients.
The missing terms in the expansion, in fact, are not generated at any order in $\alpha_s$.
The inclusion of electroweak corrections is another story: in fact, even at leading order in $\kappa_Q^2$, terms that depend on the representation of the VL quark will be generated and furthermore the expansion in $\zeta_i$ and $\xi_V$ will be affected.
Such corrections are expected to be small, certainly much smaller that the QCD ones.
These consideration apply to all the single production mechanisms here discussed.
In the present work, we will limit ourselves to compute LO cross sections, even though NLO corrections in $\alpha_s$ may be relevant.
One may be tempted to re-scale the NLO corrections from single top channels~\cite{Campbell:2009gj} to higher masses of the top quark, however this procedure is not justified here due to the presence of diagrams absent in the top case (for instance, processes mediated by the $Z$ boson).
A naive re-scaling of the NLO single top calculation would suggest that the corrections should amount to $15\div 20\%$, however a complete NLO calculation is mandatory for the extraction of reliable numbers, and we leave it for further investigation.
We used the FeynRules implementation~\cite{wwwFeynRules} to compute the coefficients in the expansion: as an example, in Tab.~\ref{tab:singletop}, we list the results for a reference mass of $M = 600$ GeV and at a 8 TeV LHC.
In the calculation, we use a 5F scheme (including the $b$ quark in the PDFs) and use the {\it CTEQ6L1} PDF set \cite{Pumplin:2002vw}.
Tables of LO coefficients for different masses can be found at the end of Appendix~\ref{app:feynrules}.

\begin{table}[tb]
\begin{center}
\begin{tabular}{l|cc|cc|c|c}
\toprule
 & $\bar{\sigma}_{Zi}^{T\bar{t}+\bar{T}t}$ & $\bar{\sigma}_{Wi}^{T\bar{t}+\bar{T}t}$ &  $\bar{\sigma}_{Wi}^{B\bar{t}+\bar{B}t}$  & $\bar{\sigma}_{Wi}^{Bt+\bar{B}\bar{t}}$ & $\bar{\sigma}_{Wi}^{X\bar{t}+\bar{X}t}$ & $\bar{\sigma}_{Wi}^{Yt+\bar{Y}\bar{t}}$\\
\midrule
$\zeta_1 = 1$ & - & 1690 & - & 3791 & 3730 & 1760 \\
$\zeta_2 = 1$ & - & 247 & - & 129 & 127 & 256 \\
$\zeta_3 = 1$ & 12.6 & 78.2 & 12.4 & - & 13.5 & 85.3 \\
\bottomrule
\end{tabular}
\caption{Coefficients for single production cross sections (in fb) in association with a top (and antitop) at 8 TeV for $M=600$ GeV. The values have been computed at LO with MadGraph.} \label{tab:singletop}
\end{center}
\end{table}

\begin{table}[tb]
\begin{center}
\begin{tabular}{l|cc|cc|c|c}
\toprule
 & $\bar{\sigma}_{Zi}^{T j+\bar{T}j}$ & $\bar{\sigma}_{Wi}^{Tj+\bar{T}j}$ &  $\bar{\sigma}_{Zi}^{B j+\bar{B}j}$  & $\bar{\sigma}_{Wi}^{B j+\bar{B} j}$ & $\bar{\sigma}_{Wi}^{X j+\bar{X} j}$ & $\bar{\sigma}_{Wi}^{Y j+\bar{Y} j}$\\
\midrule
$\zeta_1 = 1$ & 69200 & 51500 & 38100 & 62600 & 98600 & 37700 \\
$\zeta_2 = 1$ & 5380 & 10700 & 8880 & 6350 & 6490 & 10440 \\
$\zeta_3 = 1$ & - & 4230 & 3490 & - & - & 4110 \\
\bottomrule
\end{tabular}
\caption{Coefficients for single production cross sections (in fb) in association with a light jet at 8 TeV for $M=600$ GeV. The values have been computed at LO with MadGraph.} \label{tab:singlejet}
\end{center}
\end{table}

A similar expansion can be obtained for production in association with light jets (including the $b$):
\begin{eqnarray}
\sigma (Q j + \bar Q j)=\kappa_Q^2 \left(\xi_W \sum_{i=1}^3 \zeta_i\; (\bar{\sigma}^{Q jet}_{Wi} + \bar{\sigma}^{\bar Q jet}_{Wi}) + \xi_Z \sum_{i=1}^3 \zeta_i\; (\bar{\sigma}^{Q jet}_{Zi} + \bar{\sigma}^{\bar Q jet}_{Zi}) 
\right)\,,
\end{eqnarray}
where we again neglect interference terms, and the expansion can be used for NLO calculation in $\alpha_s$.
The numerical results for the coefficients $\bar{\sigma}$ are listed in Tab.~\ref{tab:singlejet}.
Generically, the production cross section with jets are much larger than with a top, and particularly large numbers are obtained for couplings to the first generation of quarks due to the valence quark enhancement.
However, the bounds on $\kappa_Q$ crucially depend on which generation the VL quark couples to, so this observation is not enough to quantify how large the single production can actually be.
We will come back to this point at the end of this section.

\subsection{Single production with bosons ($W$, $Z$ and $H$)}

\begin{table}[tb]
\begin{center}
\begin{tabular}{l|ccc|ccc|c|c}
\toprule
 & $\bar{\sigma}_{i}^{T Z+\bar{T}Z}$ & $\bar{\sigma}_{i}^{T H+\bar{T} H}$ & $\bar{\sigma}_{i}^{T W +\bar{T} W}$ &  $\bar{\sigma}_{i}^{B Z+\bar{B}Z}$  & $\bar{\sigma}_{i}^{B H+\bar{B} H}$ & $\bar{\sigma}_{i}^{B W+\bar{B} W}$ & $\bar{\sigma}_{i}^{X W+\bar{X} W}$ & $\bar{\sigma}_{i}^{Y W+\bar{Y} W}$\\
\midrule
$\zeta_1 = 1$ & 5480 & 3610 & 2430 & 2510 & 1820 & 5320 & 5320 & 2435 \\
$\zeta_2 = 1$ & 202 & 133 & 374 & 386 & 267 & 196 & 196 & 374 \\
$\zeta_3 = 1$ & - & - & 122 & 125 & 84.8 & - & - & 122 \\
\bottomrule
\end{tabular}
\caption{Coefficients for single production cross sections (in fb) with a boson at 8 TeV for $M=600$ GeV. The values have been computed at LO with MadGraph.} \label{tab:singleboson}
\end{center}
\end{table}

Here it is important to notice the presence of channels which are specific to VL quarks: production in association with $Z$ or $H$ require the presence of FCNC, absent in fourth chiral generation extensions.
Such processes, at LO, are initiated by a gluon-quark fusion. 
The relevant cross sections can be written as:
\begin{eqnarray}
\sigma (Q W^\pm + \bar{Q} W^\mp) &=& \kappa_Q^2 \left(\xi_W \sum_{i=1}^3 \zeta_i\; (\bar{\sigma}^{Q W}_{i}+\bar{\sigma}^{\bar Q W}_{i}) \right)\,, \\
\sigma (Q Z + \bar{Q} Z)         &=& \kappa_Q^2 \left(\xi_Z \sum_{i=1}^3 \zeta_i\; (\bar{\sigma}^{Q Z}_{i}+\bar{\sigma}^{\bar Q Z}_{i}) \right)\,, \\
\sigma (Q H + \bar{Q} H)         &=& \kappa_Q^2 \left(\xi_H \sum_{i=1}^3 \zeta_i\; (\bar{\sigma}^{Q h}_{i}+\bar{\sigma}^{\bar Q h}_{i}) \right)\,,
\end{eqnarray}
The above expressions are general, and extendible to NLO in $\alpha_s$.
For $X_{5/3}$ and $Y_{-4/3}$ the production in association with $Z$ or $H$ are not present due to the absence of neutral currents.
The relevant coefficients in the expansion can be found in Tab.~\ref{tab:singleboson} at LO.

\subsection{Single production in flavour-motivated benchmark models}

In Section~\ref{sect:benchmarks}, we used the simplified model with a single VL representation and bounds from flavour and electroweak precision physics to define a few benchmark points that allow to maximise the couplings in this minimal scenario.
Such benchmark models can be used to have a realistic estimate of the single production cross sections.
In the presence of a bottom partner $B$ or $Y$, flavour constraints tend to prefer a sizeable coupling with a single generation while generic couplings to the three generations would require significantly stronger bounds on the overall couplings.
For models without a bottom partner, on the other hand, sizeable couplings to two generations, one of which is the third, are allowed.

\begin{table}[tb]
\begin{center}
\begin{tabular}{lc|c|c|c|c|c|c}
\toprule
 && Benchmark 1 & Benchmark 2 & Benchmark 3 & Benchmark 4 & Benchmark 5 & Benchmark 6 \\
 && $\kappa = 0.02$ & $\kappa=0.07$ & $\kappa=0.2$ & $\kappa=0.3$ & $\kappa=0.1$ & $\kappa=0.3$ \\
 && $\zeta_1=\zeta_2=1/3$ & $\zeta_1=1$ & $\zeta_2=1$ & $\zeta_3=1$ & $\zeta_1=\zeta_3=1/2$ & $\zeta_2=\zeta_3=1/2$ \\
\midrule
\midrule
$(1,2/3)$ & $T$ & 15 (91\%) & 464 (91\%) & 564 (94\%) & 399 (95\%) & 495 (91\%) & 834 (95\%) \\
\midrule
$(1,-1/3)$ & $B$ & 14 (89\%) & 455 (88\%) & 457 (94\%) & 167 (94\%) & - & - \\
\midrule
\midrule
$(2,1/6)$ & $T$      & 5.6 (89\%) & 191 (88\%) & 114 (94\%) & 0.6 (0\%) & 195 (88\%) & 128 (94\%) \\
$\lambda_d=0$ &  $B$ & 10 (88\%) & 351 (87\%) & 267 (95\%) & 1.1 (0\%) & 358 (87\%) & 301 (95\%) \\
\midrule
$(2,1/6)$ & $T$      & 9.5 (93\%) & 272 (93\%) & 451 (94\%) & 398 (95\%) & - & - \\
$\lambda_u=0$ &  $B$ & 3.7 (90\%) & 103 (90\%) & 190 (93\%) & 166 (94\%) & - & - \\
\midrule
$(2,1/6)$ & $T$              & 15 (91\%) & 464 (91\%) & 564 (94\%) & 399 (95\%) & - & - \\
$\lambda_d=\lambda_u$ &  $B$ & 14 (89\%) & 455 (88\%) & 457 (94\%) & 167 (94\%) & - & - \\
\midrule
$(2,7/6)$ & $X$ & 15 (92\%) & 528 (92\%) & 272 (95\%) & 1.2 (0\%) & 538 (91\%) & 307 (95\%) \\
          & $T$ & 5.6 (88\%) & 191 (88\%)  & 114 (94\%) & 0.6 (0\%) & 195 (88\%) & 128 (94\%) \\
          
\midrule
$(2,-5/6)$ & $B$ & 3.7 (91\%) & 103 (90\%) & 190 (93\%) & 166 (94\%) & - & - \\
           & $Y$ & 7.6 (91\%) & 205 (90\%) & 443 (94\%) & 388 (95\%) & - & - \\
\midrule
\midrule
$(3,2/3)$ & $X$ & 30.5 (92\%) & 1055 (92\%) & 545 (95\%) & 2.4 (0\%) & - & - \\
          & $T$ & 15 (91\%) & 464 (91\%) & 564 (94\%) & 399 (95\%)   & - & - \\
          & $B$ & 7.4 (91\%) & 207 (90\%) & 380 (93\%) & 332 (94\%)  & - & - \\
                
\midrule
$(3,-1/3)$ & $T$ & 5.6 (89\%) & 191 (88\%) & 114 (94\%) & 0.6 (0\%)  & - & - \\
           & $B$ & 7.1 (89\%) & 227 (88\%) & 228 (94\%) & 84 (94\%)  & - & - \\
           & $Y$ & 7.6 (91\%) & 205 (90\%) & 443 (94\%) & 388 (95\%) & - & - \\
\bottomrule
\end{tabular}
\caption{Inclusive single production cross sections (in fb) for a single VL multiplet in various benchmark points at 8 TeV and for $M = 600$ GeV. Within brackets, the relative contribution of the processes of production in association with light quarks.} \label{tab:singlebenchmark}
\end{center}
\end{table}

    \begin{table}[tb]
    \begin{center}
    \begin{tabular}{c|cccccc}
    \toprule
     & $\kappa = 0.02$ & $\kappa=0.07$ & $\kappa=0.2$ & $\kappa=0.3$ & $\kappa=0.1$ & $\kappa=0.3$ \\
     & $\zeta_1=\zeta_2=1/3$ & $\zeta_1=1$ & $\zeta_2=1$ & $\zeta_3=1$ & $\zeta_1=\zeta_3=1/2$ & $\zeta_2=\zeta_3=1/2$ \\
    \midrule
    $\sigma_{Tq}$ & 13.8 & 422 & 533 & 380 & 451 & 790\\
    $\sigma_{T\bar t}$ & 0.3 & 8 & 10 & 8 & 9 & 15 \\
    $\sigma_{T W^-}$ & 0.4 & 12 & 15 & 11 & 13 & 22 \\
    $\sigma_{TZ}$ & 0.4 & 13 & 4 & 0 & 13 & 4 \\
    $\sigma_{TH}$ & 0.2 & 9 & 2 & 0 & 9 & 3 \\
    \midrule
    Total & 15 & 464 & 564 & 399 & 495 & 834 \\
    \bottomrule
    \end{tabular}
    \caption{Contribution of different channels to the total cross sections (in fb) for a T singlet at 8 TeV for $M = 600$ GeV.} \label{tab:relativecontributions}
    \end{center}
    \end{table}
    
\begin{table}[tb]
\begin{center}
\begin{tabular}{r|c|c|c}
\toprule
& $600$ GeV & $800$ GeV & $1000$ GeV \\
\midrule
$\begin{array}{c} (1,2/3) \\ \kappa = 0.02\quad\zeta_1=\zeta_2=1/3\end{array}$ & 15 & 7.3 & 3.8 \\
\midrule
$\begin{array}{c} (1,2/3) \\ \kappa = 0.3\quad\zeta_2=\zeta_3=1/2\end{array}$ & 834 & 324 & 138 \\
\bottomrule
\end{tabular}
\caption{Inclusive single production cross sections (in fb) for a T singlet mixing with all generations or mixing mostly with second and third generation and at 8 TeV, for 3 values of the VL mass.} \label{tab:singlebenchmark2}
\end{center}
\end{table}

In Tab.~\ref{tab:singlebenchmark} we listed the inclusive single VL quark production cross sections, computed starting from the values in Tabs~\ref{tab:singletop}, \ref{tab:singlejet} and \ref{tab:singleboson}.
While the precise numbers are just indicative, they can be used to deduce some general properties of the VL quarks.
In fact, from the second column, one can see that the case of sizeable couplings to all generation gives rise to smallish cross sections, amounting to a few fb at most, due to the strong suppression from flavour bounds.
In such a case, therefore, the searches should focus on pair production.
On the other hand, if exclusive coupling to a single generation assumed, the cross sections grow to a few hundreds fb.
Interestingly, the values of the cross sections are close independently of which generation they couple to: in most cases, the suppression due to PDF effects is compensated by a weaker flavour bound on the couplings.
There are exceptions in some cases for couplings to the third generation, mainly driven by the absence of couplings to the $W$ boson: in fact, couplings of the $Z$ to third generation leads to small cross sections as a top is always present in the final state.
Models with sizeable couplings to the third generation and only one of the two light ones are only allowed for models without $B$ and $Y$ partners, i.e. the singlet $(1,2/3)$ and doublet $(2,7/6)$, and in the case of a standard doublet $(2,1/6)$ when the mixing in the down sector is set to zero ($\lambda_d = 0$).
In all cases, large cross sections are obtained.

We emphasise the relevance of the contribution of the channels of production in association with light quarks to the total cross section: when allowed, it always contributes to around 90\% of the cross section. In Tab.\ref{tab:relativecontributions} we show in more detail the contributions of all channels for a choice of some specific scenarios.

\begin{figure}[htb]
\begin{center}
\includegraphics[width=.45\textwidth]{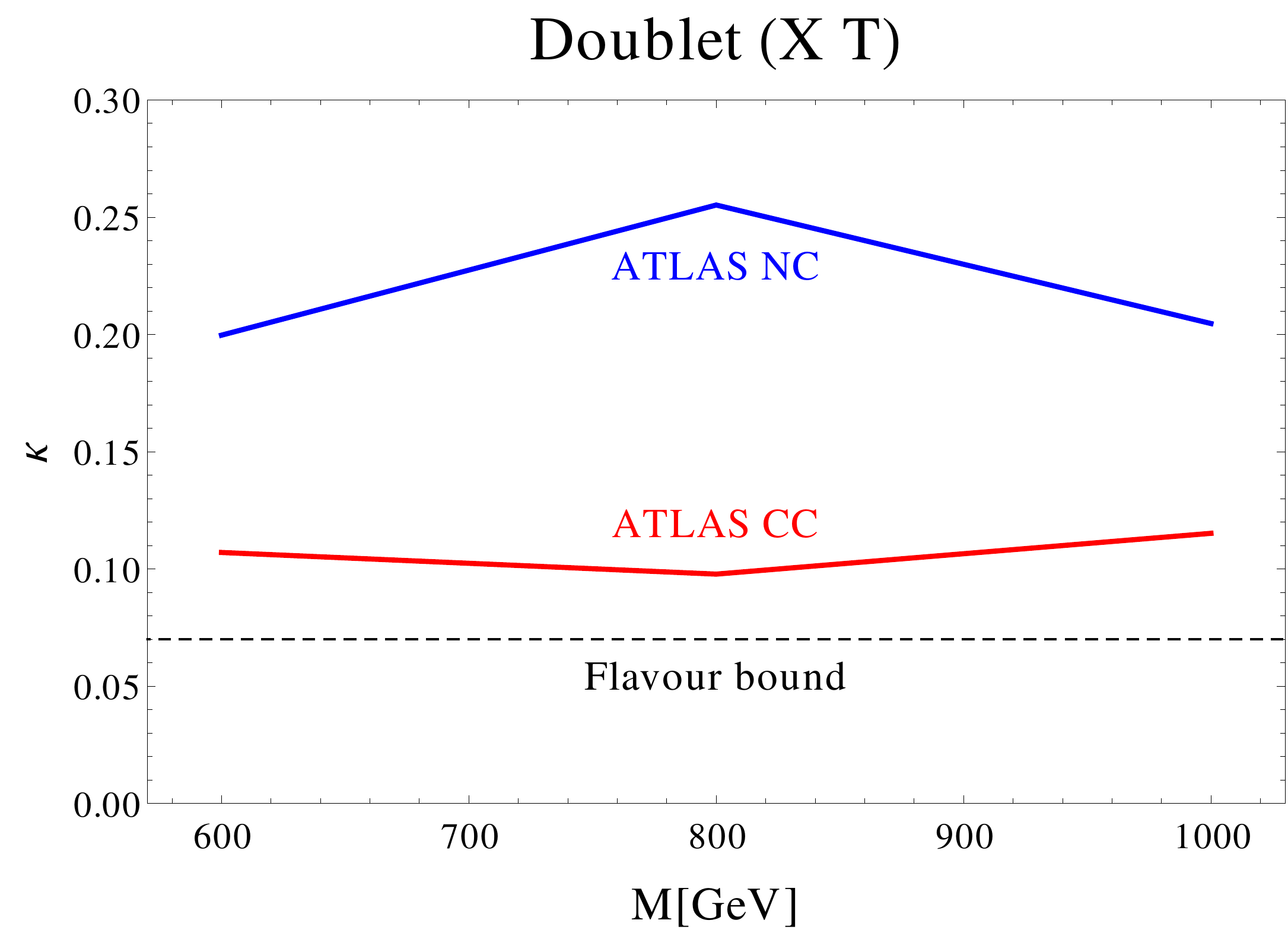}
\includegraphics[width=.45\textwidth]{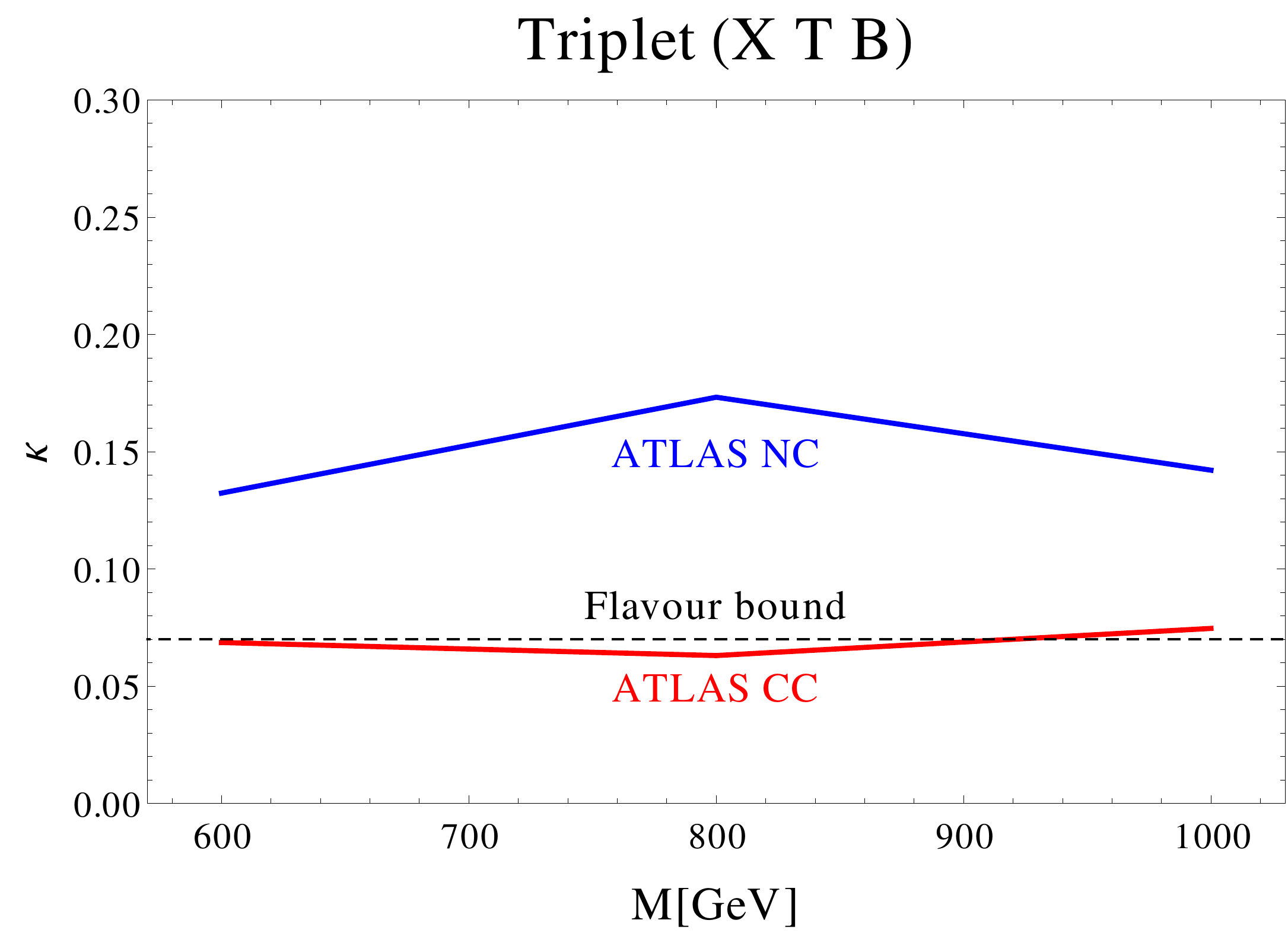}
\end{center}
\caption{Comparison of the upper bounds on the coupling strength coming from ATLAS search in single production \cite{ATLAS:2012apa} and from flavour observables considering Benchmark 2 ($\kappa=0.07$ and $\zeta_1=1$). The bound from ATLAS is competitive with the bound from flavour physics only for the triplet $(X~T~B)$ representation.} 
\label{fig:ATLASbound}
\end{figure}

The benchmark points defined in Tab.\ref{tab:singlebenchmark} can be used to investigate the sensitivity of current searches at the LHC. To date, there is only one search for single production of vector-like quarks, and it has been performed by ATLAS under the hypothesis of mixing only with light generations \cite{ATLAS:2012apa}: in Fig.\ref{fig:ATLASbound} we compare the upper bound on the coupling strength $\kappa$ obtained considering the final state investigated by ATLAS in the neutral and charged current channels with the bound coming from flavour physics observables. In order to maximise the performance of the ATLAS search we have considered the second benchmark, in which the VLQ is coupled only to the first generation. The signal in the two channels has been obtained considering all the particles in the multiplet that can contribute to the final state (e.g., in the non-SM doublet $(X~T)$ case, the CC channel receives a contribution only from the $X$, while the NC channel receives only the contribution 
of the $T$). The comparison shows that, in the range of masses considered, the bounds coming from flavour physics are usually stronger than those coming from the direct search undertaken by ATLAS and that the only representation for which the ATLAS bound is competitive with flavour bounds is the triplet $(X~T~B)$. The ATLAS analysis considers also a channel where a negative lepton is required in the final state: including this channel we obtain stronger bounds for representations that contain a $B$ or $Y$ quark, however the limits from this channel are never competitive with the flavour bounds (again, in the range of masses considered), and therefore they have not been included in the plots.

Finally, in Tab.~\ref{tab:singlebenchmark2} we show how the cross sections scale with the mass of the VL quark: we choose a representative case for illustration purposes.
We can see here that the cross sections decrease with the mass slower than the pair production ones in Tab.~\ref{tab:pairprod}, so that their relevance will be increased when higher mass regions are explored at the LHC.

\subsection{Mono production through gluon chromomagnetic coupling}

We evaluate in this section the relevance of new
chromomagnetic couplings between VL and Standard Model quarks, as the strong gluon-quark fusion
production process $qg\rightarrow Q (g)$ might not be negligible compared to the
electroweak production channels for large gluon luminosities. In this framework, $W$, $Z$ and Higgs boson loops
can induce photon and gluon couplings proportional to the very same mixing
angles that appear in Eq. (\ref{eq:param}).\ Although the loop contributions
are highly model-dependent, the $Qqg(g)$ coupling can be parametrised by%
\begin{equation}
\mathcal{L}=\frac{g_{s}g^{2}}{64\pi ^2 M} [\kappa _{gQ}\sqrt{\zeta
_{i}}(\bar{q}_{R/L}^{i}\sigma ^{\mu \nu }Q_{L/R})G_{\mu \nu }]+h.c.%
\text{ },
\end{equation}%
where $\kappa_{gQ}$ contains the details of the loop (and is proportional to the $\kappa_Q$ parameters).
In simple models containing only a single VL representation, the contribution of these loops to the branching ratios for new $T$ ($B$) quarks decaying into $gq$ are in the range $10^{-4}-10^{-6}$ for masses between $600$ and $1000$ GeV~\cite{Cacciapaglia:2010vn}. 
A similar operator with the gluon fields replaced by the photon one (and proportional to the electric charge $e$ instead of $g_s$) is also generated, and can contribute to decays $Q \to q_i \gamma$, however they are numerically smaller than the gluon one.

More in general, one can consider the chromomagnetic and electromagnetic operators as dimension-6 operators involving the new quarks.
Such operators have been listed in~\cite{Nutter:2012an,Cai:2012ji} for $T$ and $B$ partners coupling mainly to tops: in this framework, the new couplings can be parametrised in terms of the following operator suppressed by a scale $1/\Lambda^{2}$
\begin{equation}
\mathcal{L}=\kappa _{g}\frac{g_{s}v}{2\Lambda ^{2}}f_{i}^{L/R}\text{ }\bar{Q}%
_{R/L}G_{\mu \nu }\sigma ^{\mu \nu }q_{L/R}^{i}+h.c.,
\label{eq:anomalous}
\end{equation}
where $G_{\mu \nu }$ denotes the field strength tensor of the gluon, $v$ is
the Higgs field vacuum expectation value and $g_{s}$ the QCD\ coupling \cite{Degrande:2010kt}. The
parameter $\kappa _{g}$ specifies the strength of the corresponding $Qqg(g)$
strong interaction for a given VL\ quark $Q$, induced at energy scales
larger than $\Lambda $. 
A similar operator can be written for the coupling to a single photon.
In the following, the couplings $f_{i}^{L/R}\sim \sqrt{\zeta _{i}}$ will
be assumed to be of order one with the choice $\Lambda =10$ TeV. We point
out, however, that this choice is arbitrary, given that the couplings in Eq. (\ref%
{eq:anomalous}) must remain in the validity region of the perturbative
expansion. For convenience, we limit this analysis to the simplest scenario where $Q$ interacts
with the lighter families only through the above chromomagnetic
interactions. 

An operator in the form of Eq.~(\ref{eq:anomalous}) has already been considered in the context of single production of VL quarks in Ref.~\cite{Cai:2012ji}, however with a suppression scale of 1 TeV.
Here we want to keep a more conservative attitude, as such an operator is typically generated at loop level in minimal extensions of the SM, therefore we expect its coefficient to be suppressed. For instance, in the case of loops of the VL quark itself, its contribution to single production would be negligible.
The effect is also sub-leading when we consider a suppression scale around 10 TeV, as shown in Appendix~\ref{app:feynrules}.

Nevertheless, we included such operator in the FeynRules implementation of the model, with parameters that can be tuned independently on the others in Eq.~(\ref{eq:param}), so that the user can calculate its effect.
In general, such operator, which can be written for $T$ and $B$ only, will contribute to single production $g q_i \to Q_i$, single production with a jet $q_i g \to Q g$ and $g g, q \bar{q} \to Q \bar{q}_i$, and production in association with a top via a coupling to the third generation $g g, q\bar{q} \to T \bar{t}$. As shown in Tab.\ref{tab:sigmasAnomalous}, for $f_{L}^{1}=1$, the VL quarks are produced at a rate larger than anti-quarks due to the valence-sea quarks PDF
difference, while the rate is driven by the heavy quark mass in the final state for $f_{L}^{2,3}=1$. If a large coefficient for this operator is generated in models of New Physics, its effect should be added to the contribution of the other single production channels.

\section{Analysis of numerical results}
\label{sec:analysis}

\begin{figure}[htb]
\begin{center}
\includegraphics[width=.45\textwidth]{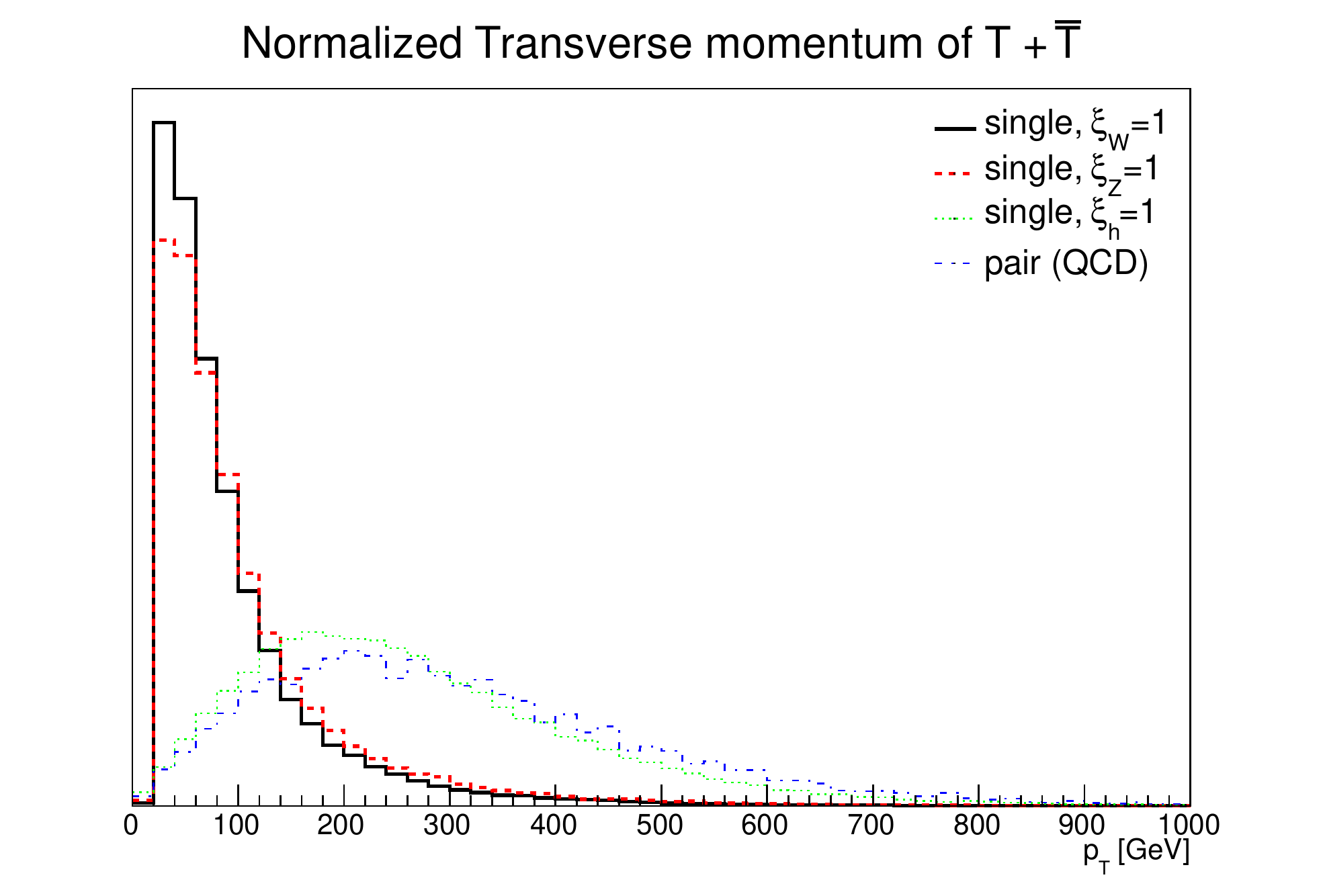}
\includegraphics[width=.45\textwidth]{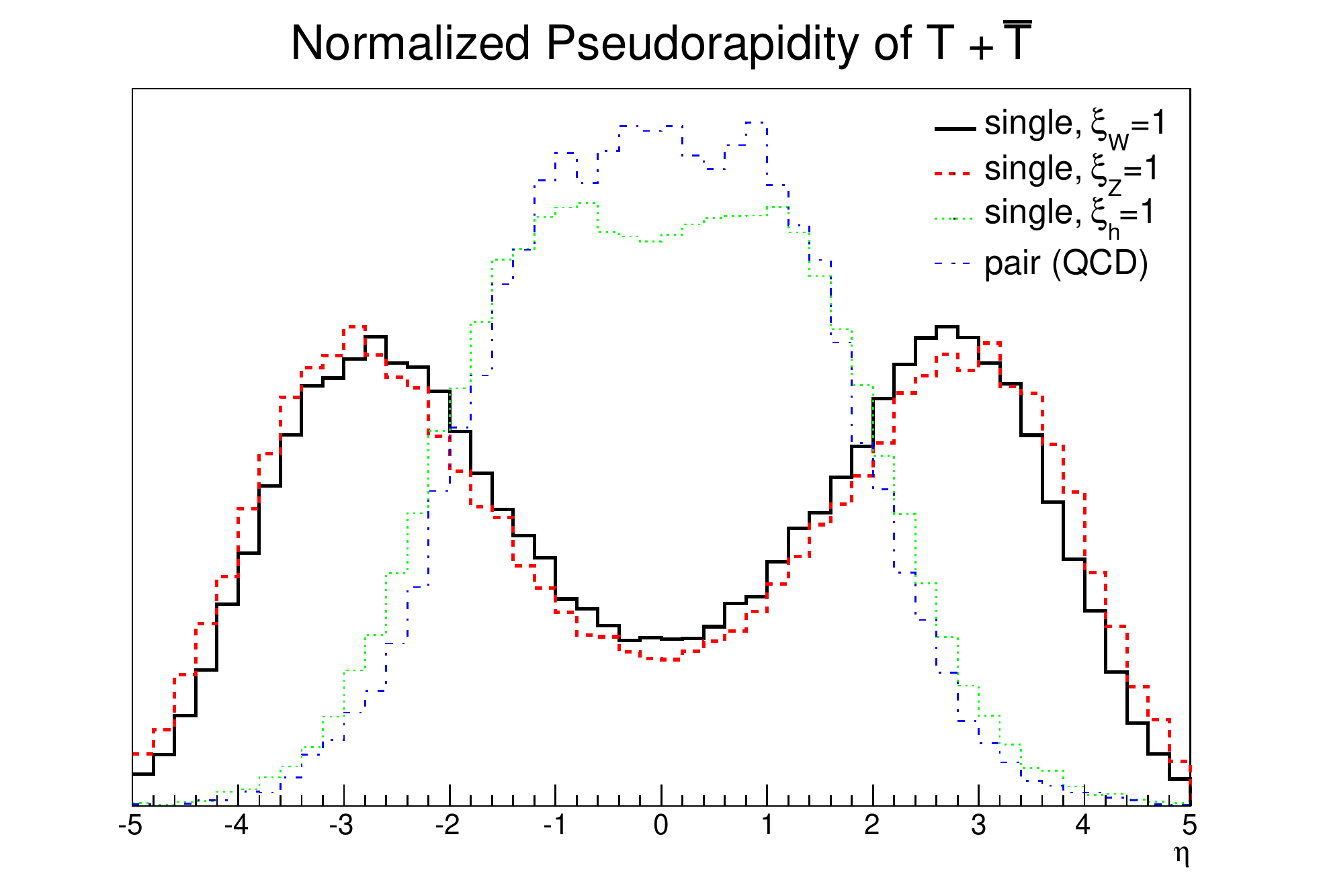}\\
\includegraphics[width=.45\textwidth]{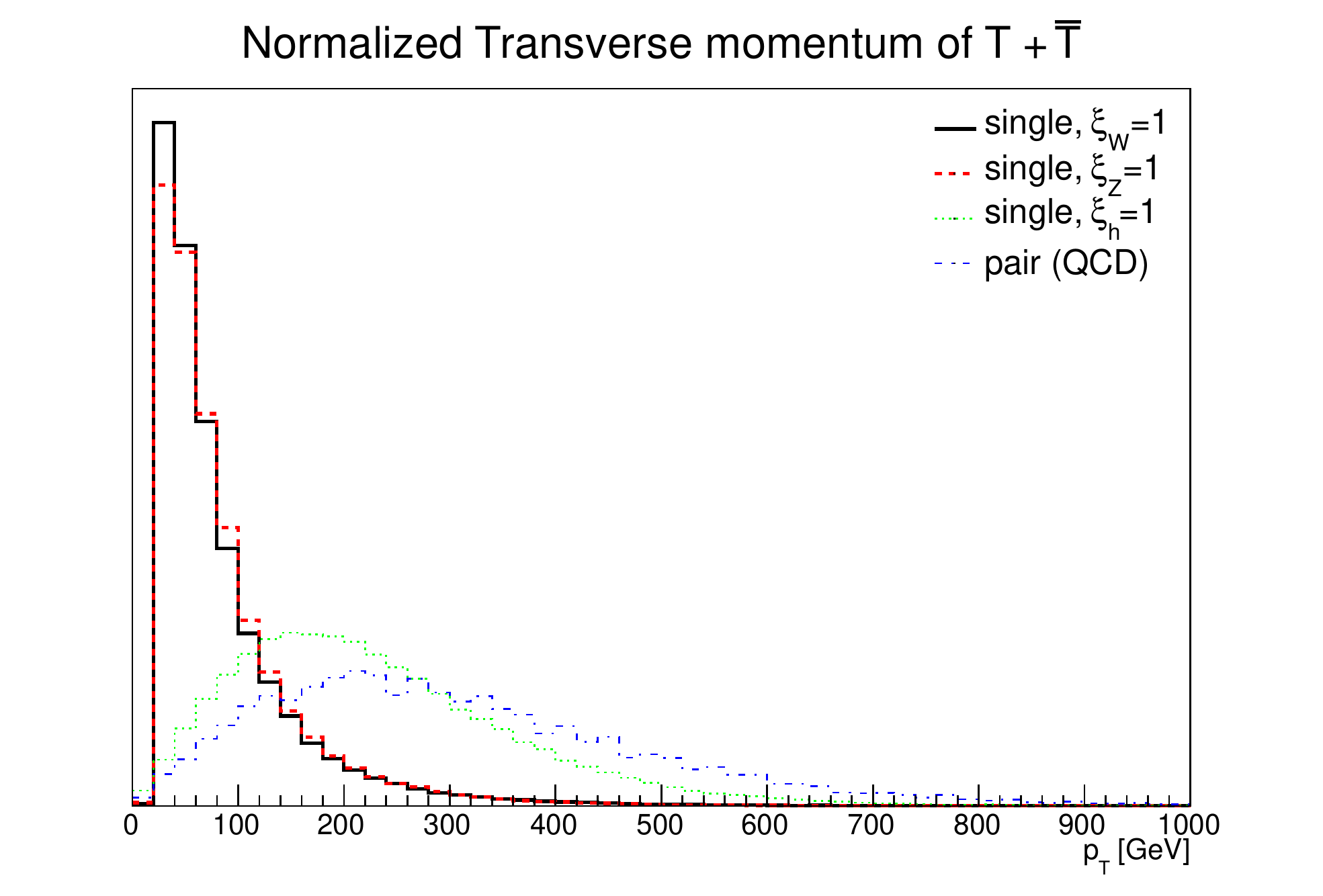}
\includegraphics[width=.45\textwidth]{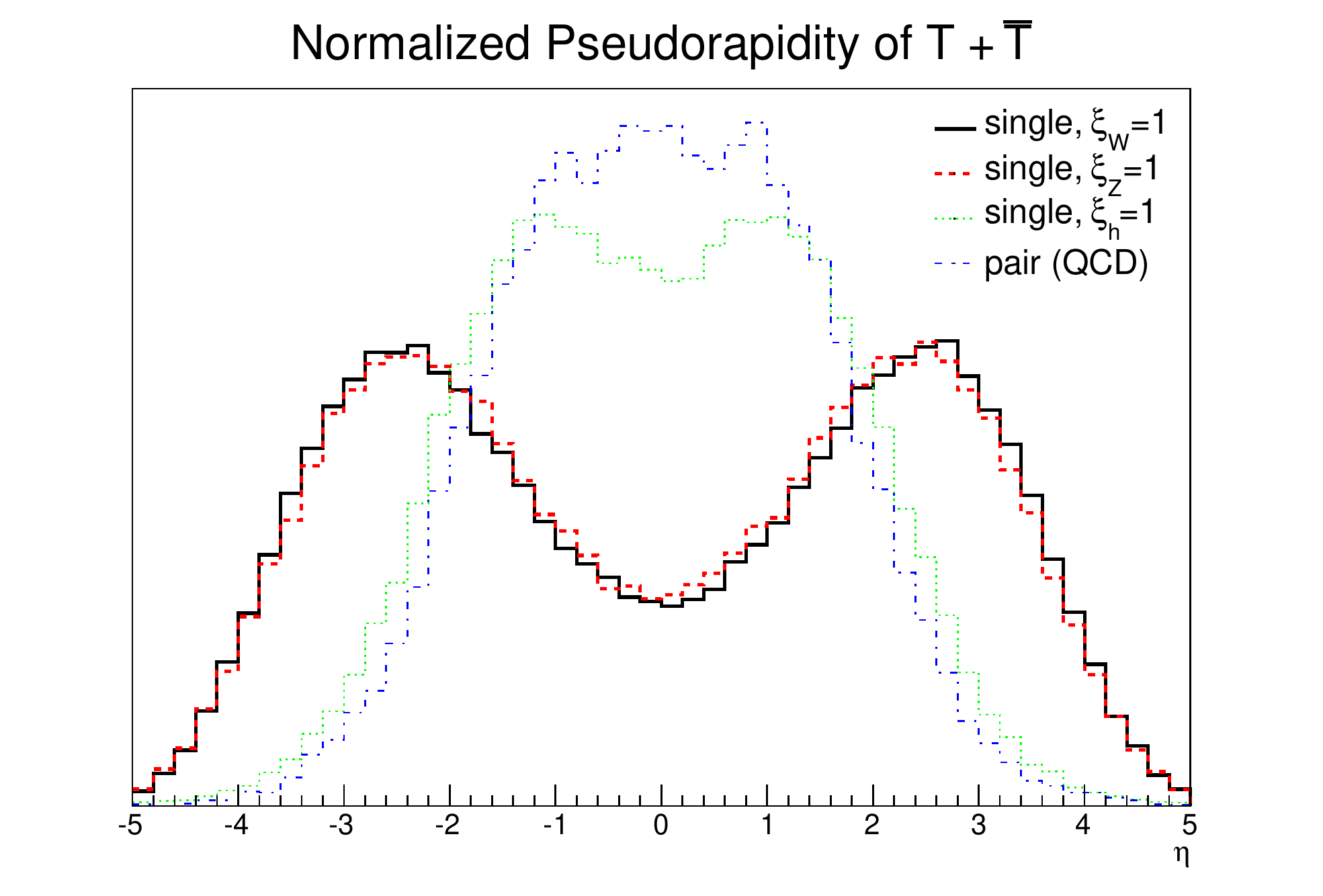}\\
\includegraphics[width=.45\textwidth]{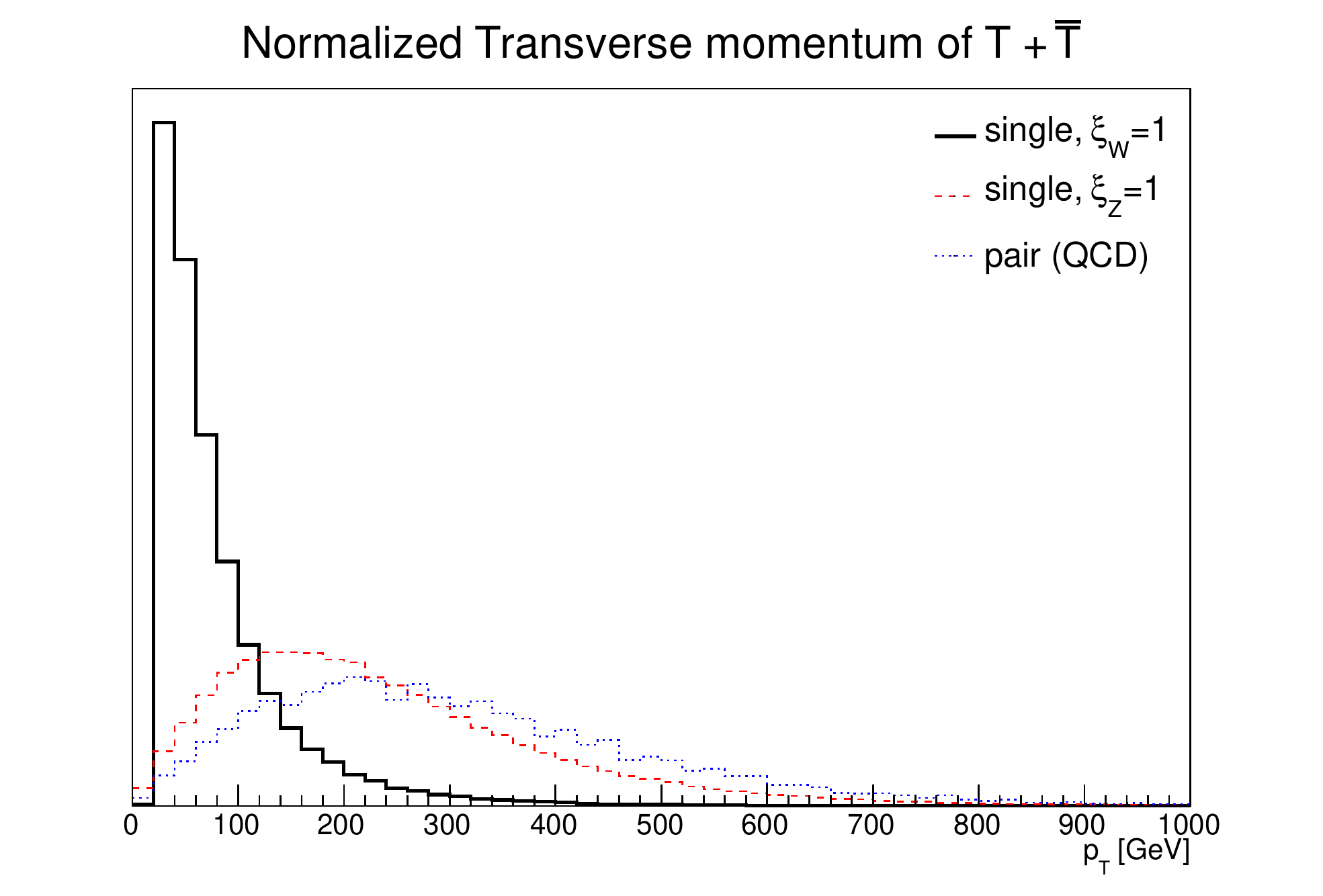}
\includegraphics[width=.45\textwidth]{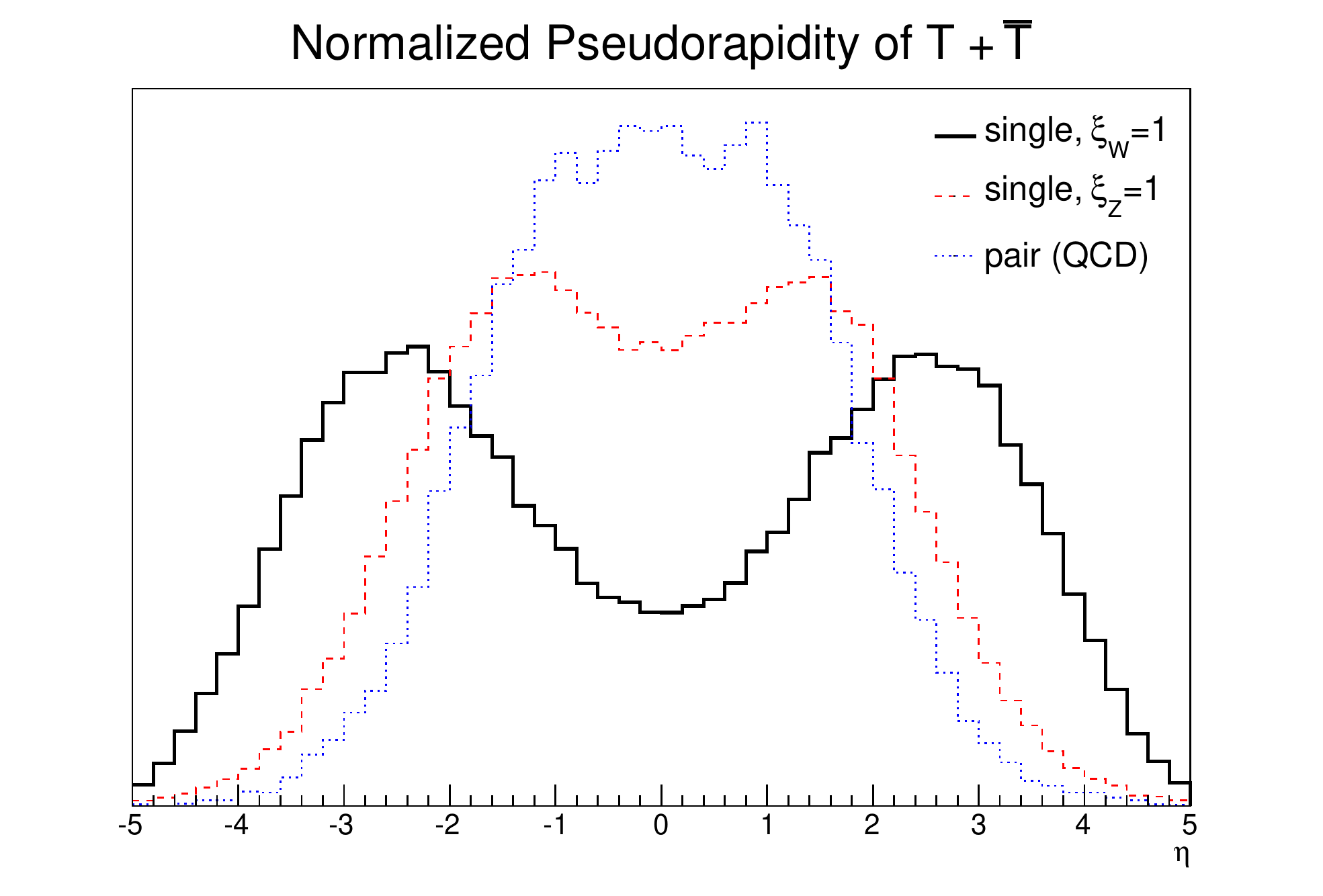}
\end{center}
\caption{Normalised transverse momentum and pseudorapidity distributions for inclusive single production of $T$ compared to QCD pair production with, from top to bottom, $\zeta_1=1$, $\zeta_2=1$, $\zeta_3=1$.} 
\label{fig:plotssamekin}
\end{figure}

The analysis performed so far allows us to highlight some main conclusions that can be useful to drive searches of
VL quark production in a model-independent fashion. The main messages we would like to convey are the following:

\begin{itemize}
\item \textit{Relevance of single production}: new top partners can have sizeable production rates regardless of their mixing structure with
first, second or third generation quarks without conflicting with current experimental constraints. This is due to a compensation between suppression coming from PDFs and from flavour observables, as shown in Table~\ref{tab:singlebenchmark}. 
Processes like, for instance, production with a light jet followed by decays into third generation should therefore be considered.
\item \textit{Exclusive mixing hypotheses}: assuming exclusive (100\%) branching ratios may forbid some single production channels. 
In the case of exclusive mixing with third generation, the channels $Xj,XW,TZ,TH$ and $BW$ are systematically forbidden. Therefore, considering a scenario with $T$ mixing only with third generation, the Higgs and $Z$ bosons may only appear in the $T$ decay products.
Analogous conclusion can be derived for the presence of a $W$ boson in scenarios with $B$ or $X$ partners, coupling only to third generation.
Furthermore, $Bt$ and $Bj$ ($Tj$) production are not allowed in the latter case if $\xi_{W}=1$ ($\xi _{Z}=1$), whereas $pp\to TH,TZ,BW,XW$ can only arise in scenarios where the VLQ mixes with either the first or second generation. 
The complete list of forbidden channels under different assumptions on mixings is in Tab.\ref{tab:forbiddenchannels}:
\begin{table}[tb]
\begin{center}
\begin{tabular}{ccc}
\toprule 
          & $\zeta_{1,2}=1$ & $\zeta_3=1$ \\
\midrule
$\xi_W=1$ & $\begin{array}{c} TZ,TH \\ B\bar{t},BZ,BH \end{array}$ & $\begin{array}{c} Xj,XW \\ TZ,TH \\ Bj,B\bar t,BW,BZ,BH\end{array}$\\
\midrule
$\xi_Z=1$ & $\begin{array}{c} T\bar{t},TW,TH \\ Bt,B\bar{t},BW,BH \end{array}$ & $\begin{array}{c} Tj,TW,TZ,TH \\ Bt,B\bar t,BW,BH\end{array}$\\
\midrule
$\xi_H=1$ & $\begin{array}{c} \mbox{all channels but } TH \mbox{ are forbidden} \\ \mbox{all channels but } BH \mbox{ are forbidden} \end{array}$ & $\begin{array}{c} \mbox{all channels are forbidden} \\ \mbox{all channels but } BH \mbox{ are forbidden} \end{array}$\\
\bottomrule
\end{tabular}
\caption{Forbidden channels for single production under hypothesis of exclusive (100\%) mixing patterns.} \label{tab:forbiddenchannels}
\end{center}
\end{table}

\item \textit{Associated production with top quarks}: VL quark single production in association with a top (antitop) quark provides a very
interesting final state for the forthcoming searches. As shown in Tab.\ref{tab:relativecontributions}, $pp \to Qt$ is worth exploring even in scenarios where the VL quark does not mix exclusively with the third family, due to the PDF enhancement in production.

\item \textit{Distributions:} for $\zeta _{1,2}=1$, the transverse momentum $p_{T}$ and rapidity $\eta$
distributions for inclusive $T$ and $B$ production, 
distinguishing amplitudes from $W$ and $Z$ exchanges,
are hardly distinguishable at the level of production, as can be seen in Fig.\ref{fig:plotssamekin}, where distributions related to $T$ have been considered for illustrative purpose. 
We can therefore infer that the production does not distinguish between couplings to first and second generation (while the value of the cross sections does depend on this), so that at the level of generation one can consider either one of them.
Furthermore, no kinematical cuts should be able to distinguish the two.
We will confirm these observations with a detailed simulation that includes the kinematics of the decays.

\item \textit{Electro-weak $QQ^{(\prime)}$ production}: this channel is suppressed by the mixing to the light quarks, nevertheless it leads to sizeable cross sections and interesting final states for unsuppressed $\kappa_Q$ scenarios. We have checked numerically that in the flavour-motivated benchmarks defined in the previous sections, the cross sections for EW $QQ^{(\prime)}$ production are below the femtobarn in almost all cases, making them negligible. However, they should be included in studies of single production focused to non-minimal models where $\kappa_Q$ is not suppressed and if large mixing to first generation is allowed.

\end{itemize}

\section{Conclusions and outlook}
\label{sec:conclusion}

In this work we have developed a model-independent parametrisation that can be used to describe the phenomenology of vector-like 
quarks with generic hypotheses about their mixing with SM quarks. The framework relies on a limited number of parameters which
represent all possible mixings with SM quarks and couplings with SM bosons. The parametrisation is implemented in a 
publicly available FeynRules model~\cite{wwwFeynRules}, and has been adopted to perform an analysis of processes of production of vector-like quarks: 
single production, mono production through chromomagnetic coupling, off-diagonal pair production ($\bar{Q}Q^\prime$) and same-sign EW 
pair production ($QQ^{(\prime)}$). The main result of the present analysis is to provide a theoretically consistent and general framework, where the various searches for VL or fourth generation quarks can be embedded in.
In particular, our framework allows for a simple correlation of the branching ratios with the single production channels.
We also provided a LO calculation of all the relevant single production cross sections, via model independent coefficients.

To provide a concrete application of the parametrisation, we have applied this technique to obtain the production cross sections for some benchmark points which satisfy experimental constraints from flavour physics and other observables. Our analysis, even if limited to production level, allows to highlight the potential relevance of scenarios which 
have been neglected in previous experimental searches, such as mixing only with second generation or presence of tops in the final 
state in scenarios where the vector-like quark does not mix with third generation at all. These points deserve special attention as they 
may be the key to obtain hints of these vector-like states in realistic flavour mixing scenarios which go beyond the na\"ive
expectations based on simplified single generation mixing. 
For instance, we showed that:
\begin{itemize}
\item single production followed by decays to third generation should be considered, as the production rates are similar to the ones obtained through couplings to the first generation;
\item single production with tops is relevant also in the case of decays into light generations;
\item depending on the hypothesis on the branching ratios, some single production channels are forbidden, at leading order in the electroweak couplings.
\end{itemize}
From the parton level simulations we performed in this work it is not possible to determine if the final states in single production are sensitive to the generation the VL quark mainly couples to, even though our distributions tend to show a mild dependence on the couplings.
A more thoroughly analysis will be performed also to design a way to disentangle pair production, which is QCD dominated and thus only depends on the VL mass, and single production, which is directly proportional to the mixing of the VL quark to standard ones.
A measure of the coupling is indeed essential to determine the nature of the VL quarks and, maybe, have a glimpse on the origin of the standard generations.

\section*{Acknowledgements}
We would like to thank Fabio Maltoni and Benjamin Fuks for useful discussions and suggestions at the early stages of this work.
This work was partially supported by CF-Theorie IN2P3. G.C. and A.D. thank ICTP-SAIFR for hospitality during the final stage of this work. The work of M.B. is supported by the National Fund for Scientific Research (F.R.S.-FNRS, Belgium) under a FRIA grant. This project has been achieved in the context of the LabEx ``Lyon Institute of Origins'' (LIO).

\appendix

\section{Vector-like quarks have chiral couplings to SM quarks}
\label{app:multiVL}

In this appendix we will study the structure of the couplings of the VL quarks to SM quarks and gauge bosons, relevant for decays and single production, in the case of an arbitrary number of VL representations.
We will show that, even in presence of mixing between various VL representations, the couplings are dominantly chiral like in the case of a single VL multiplet: this calculation agrees with the results in~\cite{delAguila:2000rc}.
In all models, such couplings can be traced back to Yukawa couplings connecting the VL multiplets with a SM chiral fermion via the Higgs boson.
This is true also for models with extended Higgs sector: a singlet which acquires a VEV will generate masses in the form of VL masses as it does not break the gauge symmetries; additional doublets will generate the same structures as the SM Higgs, and only the coupling of the Higgs boson may be affected; for larger representation, like a triplet, the VEV will generally induce large corrections to the $\rho$ parameter, thus it is bound to be very small and therefore generates small mixing terms. 
Following the above arguments, we can state that our assumption of mixing mainly via the SM Higgs is solid.
The Higgs in the SM is a doublet of SU(2), thus a field with weak isospin-1/2: a convenient way to classify the VL quarks is to use their weak isospin.
In fact, VL multiplets with integer isospin (singlets, triplets, ...) can only couple via the Higgs to a left-handed doublet; on the other hand, VL multiplets with semi-integer isospin (doublets, quadruplets...) can only couple to a SM right-handed singlet.
Also, one can potentially write a Yukawa coupling of a Higgs field with two VL quarks only if one of them has integer isospin (thus belonging to the first class) and the other semi-integer isospin.
Of course, not all couplings are allowed as one needs to take into account the specific representation and the hypercharge.
In any case, this classification allows us to write down the most general possible mass matrix which will determine the mixing between the SM quarks and the VL quarks.
As shown in Eq.(\ref{eq:Zcoups}), the FCNCs and the couplings between VL quarks and the standard ones are all proportional to the elements of the mixing matrices $V_{L/R}^{\alpha i}$, where $\alpha$ spans over the VL quarks, and $i=1,2,3$ on the SM quarks.
In the following, we will therefore focus on these elements of the mixing matrices.

Let us consider the most general case with $N-3$ VL quarks that mix via Yukawa interactions to the SM quarks, and to each other.
In the starting basis, we consider that the SM Yukawa matrices are already diagonal (for simplicity), while the VL masses are also diagonal.
We consider $n_d$ semi-integer isospin states (doublets, ...) with potential mixing with the SM right-handed singlets, and $n_s = N-3-n_d$ integer isospin states (singlets, triplets, ...) with potential mixing with the SM left-handed doublets.
The most general mass matrix, therefore, will have the following block form:
\beq
\mathcal{L}_{\rm mass} = \bar{q}_L \cdot \left( 
\begin{array}{ccc|ccc|ccc}
\mu_1 & 0 & 0 & 0 & \dots & 0 & x_{1, n_d+4} & \dots &  x_{1, N}\\
0 & \mu_2 & 0 & 0 & \dots & 0 & x_{2, n_d+4} & \dots &  x_{2, N}\\
0 & 0 & \mu_3 & 0 & \dots & 0 & x_{3,n_d+4} & \dots &  x_{3, N}\\
\hline
y_{4,1} & y_{4,2} & y_{4,3} & M_4 & 0 & 0 & & &  \\
\vdots & \vdots & \vdots & 0 & \ddots & 0  & & \omega_{\alpha \beta} & \\
y_{n_d+3,1} & y_{n_d+3,2} & y_{n_d+3,3} & 0 & 0 & M_{n_d+3} & & \\
\hline
0 & 0 & 0 & & & & M_{n_d+4} & 0 & 0 \\
\vdots & \vdots & \vdots & & \omega'_{\alpha \beta} & & 0 & \ddots & 0 \\
0 & 0 & 0 & & & & 0 & 0 & M_{N}
\end{array} \right) \cdot q_R + h.c.
\eeq
In this basis, the SM Yukawa masses $\mu_i$ are presented in a diagonalised form.
We recognise the $3\times n_s$ matrix $x_{i,\beta_s}$ of the Yukawa couplings of the VL singlets/triplets (integer isospin), and the $n_d \times 3$ matrix $y_{\alpha_d, j}$ of the Yukawa couplings of the VL doublets (semi-integer isospin).
$M_\alpha$ represent the VL masses of all the new representations, while the $n_d \times n_s$ matrix  $\omega_{\alpha_d, \beta_s}$ and $n_s \times n_d$ matrix $\omega'_{\alpha_s, \beta_d}$ contain the eventual Yukawa couplings among VL representations.
Note here that in general the Yukawa couplings between VL quarks distinguish between the chiral components of the VL quarks, therefore in general $\omega' \neq \omega^T$: furthermore, $\omega'$ corresponds to the ``wrong'' Yukawa couplings, in the sense that it connects left-handed singlets (integer isospin) with right-handed doublets (semi-integer isospin), which is the opposite chirality configuration of SM Yukawa couplings.
This fact will force $\omega'$ to play a crucial role in the following discussion.

To simplify the discussion, in the following we will neglect the SM quark masses, thus $\mu_i = 0$, and assume that all the Yukawa couplings of the VL quarks are of the same order $x \sim y \sim \omega \sim \omega' \sim v$.
The latter assumptions are only used to obtain a consistent expansion of the mixing angles.
Finally, we assume that the mass scale for the VL quark masses is larger that the Higgs VEV $v$, so that we can analyse the mixing in an expansion for small $v/M_{VL}$.
One however needs to pay extra care with the mixing between VL quarks: in fact, the mixing can be large if the difference between two VL masses is smaller than the Yukawa contribution $\omega$ and $\omega'$.
For this reason, we will diagonalise exactly the $N-3 \times N-3$ mass matrix in the VL block, defining:
\beq \label{eq:matVL}
U_L^\dagger \cdot \left( 
\begin{array}{cccccc}
M_4 & 0 & 0 & & &  \\
 0 & \ddots & 0  & & \omega_{\alpha \beta} & \\
0 & 0 & M_{n_d+3} & & \\
 & & & M_{n_d+4} & 0 & 0 \\
 & \omega'_{\alpha \beta} & & 0 & \ddots & 0 \\
 & & & 0 & 0 & M_{N}
\end{array} \right) \cdot U_R = \left( \begin{array}{ccc}
\bar{M}_4 & 0 & 0 \\
0 & \ddots & 0 \\
0 & 0 & \bar{M}_{N}
\end{array} \right)\,.
\eeq
Here $U_{L/R}$ represent the diagonalisation matrices, and $\bar{M}_\alpha$ the mass eigenvalues.
The mixing angles in $U_{L/R}$ are generally large. 
In the new basis, the mass matrix reads:
\beq
\tilde M = \left( 
\begin{array}{cccccc}
\mu_1 & 0 & 0 & a_{1, 4} & \dots &  a_{1, N}\\
0 & \mu_2 & 0 & a_{2, 4} & \dots &  a_{2, N}\\
0 & 0 & \mu_3 & a_{3,4} & \dots &  a_{3, N}\\
b_{4,1} & b_{4,2} & b_{4,3} & \bar{M}_4 & 0 & 0 \\
\vdots & \vdots & \vdots & 0 & \ddots & 0 \\
b_{N,1} & b_{N,2} & b_{N,3} & 0 & 0 & \bar{M}_{N}
\end{array} \right)\,,
\eeq
where
\beq
a_{i, \beta} = \sum_{\gamma_s=n_d+4}^{N}\; x_{i, \gamma_s} (U_R)_{\gamma_s \beta}\,, \quad b_{\alpha, j} = \sum_{\gamma_d=4}^{n_d+3}\; (U^\dagger_L)_{\alpha \gamma_d} y_{\gamma_d, j}\,,
\eeq 
are $3\times N-3$ and $N-3 \times 3$ matrices containing the information about the Yukawa couplings connecting VL quarks with the SM ones.
Now we can safely expand for small $\epsilon = v/M \sim a/M \sim b/M$, and define the mixing matrices as:
\beq
\tilde{V}_L^\dagger \cdot \tilde{M} \cdot \tilde{V}_R = M^{\rm diag}\,.
\eeq
The mass eigenstates are given by:
\beq
m^2_i &\sim& \mu_i^2\sim 0\,, \\
m^2_\alpha & \sim & \bar{M}^2_\alpha \left( 1 + \sum_{j=1}^3 \frac{(a^\dagger_{\alpha, j} a_{j, \alpha} + b_{\alpha, j} b^
\dagger_{j, \alpha} ) }{\bar{M}_\alpha^2}  + \mathcal{O} (\epsilon^3) \right)\,,
\eeq
where $i = 1, 2, 3$ and $\alpha = 4, \dots, N$.
The relevant mixing elements are given by:
\beq
\tilde{V}_L^{\alpha i} & \sim & - \frac{a^\dagger_{\alpha, i}}{\bar{M}_\alpha} + \mathcal{O} (\epsilon^3)\,, \\
\tilde{V}_R^{\alpha i} & \sim &  - \frac{b_{\alpha, i}}{\bar{M}_\alpha} + \mathcal{O} (\epsilon^3)\,,
\eeq
where we neglected terms suppressed by the SM Yukawa couplings.
The mixing matrices in the starting basis can be obtained as:
\beq
V_{L/R} = \left( \begin{array}{cc}
1 & 0 \\ 0 & U_{L/R} \end{array} \right) \cdot \tilde{V}_{L/R}\,;
\eeq
therefore
\beq
V_L^{\alpha i} &=& ( U_L \cdot \tilde{V}_L)^{\alpha i} = - \sum_{\beta=4}^N \sum_{\gamma=n_d+4}^N \frac{U_L^{\alpha \beta} (U_R^\dagger)^{\beta \gamma}}{\bar{M}^\beta} (x^\dagger)_{\gamma i}\,, \\
V_R^{\alpha i} &=& ( U_R \cdot \tilde{V}_R)^{\alpha i} = - \sum_{\beta=4}^N \sum_{\gamma=4}^{n_d+3} \frac{U_R^{\alpha \beta} (U_L^\dagger)^{\beta \gamma}}{\bar{M}^\beta} y_{\gamma i}\,.
\eeq
The mixing matrices $U_{L/R}$ can be eliminated by observing that
\beq
 \sum_{\rho=4}^{N+3}\; \frac{(U_R)_{\alpha \rho} (U^\dagger_L)_{\rho \beta}}{\bar{M}_\rho} = \left( U_R \cdot \bar{M}^{-1} \cdot U_L^\dagger \right)_{\alpha \beta} = \left( M_{VL}^{-1} \right)_{\alpha \beta}\,,
\eeq
where $M_{VL}$ is the $N-3 \times N-3$ block mass of the VL quarks in Eq.~\ref{eq:matVL}, and $\bar{M}$ the diagonalised matrix.
Finally we obtain
\beq
V_L^{\alpha i} = - \left( x \cdot M_{VL}^{-1} \right)^\dagger_{\alpha i} + \dots \,, \quad
V_R^{\alpha i} = - \left( M_{VL}^{-1} \cdot y \right)_{\alpha i} + \dots
\eeq
At first sight, for every VL quark $\alpha$, there is a sizeable mixing angle both on the left and right-handed sectors, and they arise at the same order in the expansion, however a closer look will show that this is not the case.
In fact, the matrices containing the Yukawa couplings are not $N-3 \times N-3$ matrices, but only spans over either the semi-integer or integer isospin VL quarks.
Furthermore, at leading order in $\epsilon$, the inverse VL mass matrix is given by:
\beq
M_{VL}^{-1} \sim \left( \begin{array}{cc}
M_d^{-1} & - M_d^{-1} \cdot \omega \cdot M_s^{-1} \\
- M_s^{-1} \cdot \omega' \cdot M_d^{-1} & M_s^{-1}
\end{array} \right)\,,
\eeq
where $M_d$ is the $n_d \times n_d$ diagonal matrix for the semi-integer isospin VL quarks, and $M_s$ the $n_s \times n_s$ diagonal mass matrix for the integer isospin ones.
Plugging this formula in the expression for the mixing angles, at leading order in $\epsilon$, we obtain:
\beq
\mbox{semi-integer isospin:} & \Rightarrow & \left\{ \begin{array}{l}
V_L^{\alpha_d i} \sim \left( x \cdot M_s^{-1} \cdot \omega' \cdot M_d^{-1} \right)^\dagger_{\alpha_d i} \sim \mathcal{O} (\epsilon^2) \\
V_R^{\alpha_d i} \sim - \left(M_d^{-1} \cdot y\right)_{\alpha_d i} \sim \mathcal{O} (\epsilon) 
\end{array} \right. \\
\mbox{integer isospin:} & \Rightarrow & \left\{ \begin{array}{l}
V_L^{\alpha_s i} \sim - \left(x \cdot M_s^{-1}\right)^\dagger_{\alpha_s i} \sim \mathcal{O} (\epsilon)  \\
V_R^{\alpha_s i} \sim \left(M_s^{-1} \cdot \omega' \cdot M_d^{-1} \cdot y \right)_{\alpha_d i} \sim \mathcal{O} (\epsilon^2) 
\end{array} \right.
\eeq
We can therefore see that for the semi-integer isospin (doublet) VL quarks, the dominant mixing angle is right-handed, while the left-handed one is suppressed by an extra factor of $\omega'/M_{VL}$; the chiralities are exchanged for the integer isospin (singlets and triplets) VL quarks.
It is interesting to stress that the subleading mixing angle is suppressed either by the ``wrong'' Yukawa couplings between VL quarks, $\omega'$, or by the light quark masses.
This finally proves that the couplings of VL quarks to SM ones are chiral also in the most general scenario, and that the chirality of the dominant mixing only depends on the representation the VL quark belongs to.

Note also that the leading mixing angle may be small in some cases, either because the relevant Yukawa coupling is absent or numerically small. 
However, in such a case, the couplings will receive an extra suppression and the single production will become subdominant with respect to the pair production: for the pair production studies, only the branching ratios are relevant, independently on the chirality of the couplings, so that our parametrisation remains useful in such a case too.

\section{Top mass corrections to the widths}
\label{app:topmass}

In this paper we consistently neglected the masses of the standard quarks to simplify the formulae for the branching ratios.
This is an excellent approximation, except for the top.
In this Appendix, we want to discuss the numerical impact of the corrections due to the large top mass which are however model dependent.
In fact, the top mass can enter the decay widths in two ways:
\begin{enumerate}
\item[-] modify the kinematics of the decay, thus affecting the decay widths of any final state containing a top;
\item[-] via the sub-leading coupling, which is suppressed by a factor $m_t/M$ with respect to the leading chirality one, however this contribution is model dependent as it crucially depends on the representation of the VL quarks.
\end{enumerate}
To be concrete, we will use the simple cases of a single representation in order to evaluate the impact of the corrections.
The main assumption here, which is quite general, is that the sub-leading coupling is proportional to the same flavour mixing angle as the leading one, so that the $\zeta_i$ dependence is not affected.
The only channels that are affected are $T \to Ht$, $Zt$, $W^+b$, $B \to W^-t$ and $X \to W^+t$.
The effect can be included by means of 5 model dependent functions of the $m_t/M$ ratio, so that the branching ratios of $T$, $B$ and $X$ are modified as follows:
\beq
BR(T \to W^+ j) = \frac{\zeta_{jet} \xi^T_W}{1+(1-\zeta_{jet}) \Delta_T}\,, & & BR(T \to W^+ b) = \frac{(1-\zeta_{jet}) \xi^T_W (1 + \delta_W)}{1+(1-\zeta_{jet}) \Delta_T}\,,  \nonumber\\
BR (T \to Z j) = \frac{\zeta_{jet} \xi^T_Z}{1+(1-\zeta_{jet}) \Delta_T}\,, & & BR (T \to Z t) = \frac{(1-\zeta_{jet}) \xi^T_Z (1+\delta_Z)}{1+(1-\zeta_{jet}) \Delta_T}\,,\nonumber\\
BR(T \to H j) =\frac{ \zeta_{jet} (1-\xi^T_Z-\xi^T_W)}{1+(1-\zeta_{jet}) \Delta_T}\,, & &  BR(T \to H t) = \frac{(1-\zeta_{jet})(1-\xi^T_Z-\xi^T_W)(1+\delta_H)}{1+(1-\zeta_{jet}) \Delta_T}\,, \nonumber
\eeq
where $\Delta_T = (1-\xi^T_W-\xi^T_Z) \delta_H + \xi^T_Z \delta_Z + \xi^T_W \delta_W$, and
\beq
BR(B \to W^- j) = \frac{\zeta_{jet} \xi^B_W}{1+(1-\zeta_{jet}) \xi_W^B \delta_B}\,, & & BR(B \to W^- t) = \frac{(1-\zeta_{jet}) \xi^B_W (1+\delta_B)}{1+(1-\zeta_{jet}) \xi_W^B \delta_B}\,, \nonumber \\
BR (B \to Z j) = \frac{\zeta_{jet} \xi^B_Z}{1+(1-\zeta_{jet}) \xi_W^B \delta_B}\,, & & BR (B \to Z b) = \frac{(1-\zeta_{jet}) \xi^B_Z}{1+(1-\zeta_{jet}) \xi_W^B \delta_B}\,,  \nonumber \\
BR(B \to H j) = \frac{\zeta_{jet} (1-\xi^B_Z-\xi^B_W)}{1+(1-\zeta_{jet}) \xi_W^B \delta_B}\,, & &  BR(B \to H b) = \frac{(1-\zeta_{jet})(1-\xi^B_Z-\xi^B_W)}{1+(1-\zeta_{jet}) \xi_W^B \delta_B}\,,\nonumber \\
 & & \nonumber \\
BR(X \to W^+ j) =\frac{\zeta_{jet}}{1+(1-\zeta_{jet}) \delta_X}\,, & & BR(X \to W^+ t) = \frac{(1-\zeta_{jet}) (1+\delta_X)}{1+(1-\zeta_{jet}) \delta_X}\,. \nonumber
\eeq
To be concrete and have a numerical estimate of the effects, we will calculate them explicitly in the cases of a single VL representation, discussed in Section~\ref{sect:benchmarks}. We will now discuss the 5 corrections one by one.

\subsection*{Higgs: $T \to H t$}

For the couplings of the top to the Higgs, the sub-leading vertex is always present because it originates from the mass mixing itself: we can parametrise as the leading one times a factor $c_H m_T/M$.
For a single VL representation, we find that $c_H = 1$ for all representations, so that the corrections is effectively model-independent.
The correction, therefore, is given by:
\beq
\delta_H = \sqrt{\lambda (1,\epsilon_H, \epsilon_t)} \frac{ (1+\epsilon_t-\epsilon_H)(1+c_H^2 \epsilon_t) + 4 c_H \epsilon_t }{\left( 1-\epsilon_H\right)^2} -1 \sim c_H (c_H + 4) \epsilon_t\,,
\eeq
where $\epsilon_t = m_t^2/M^2$, $\epsilon_H = m_H^2/M^2$, and we have expanded the result at leading order in $1/M^2$.
For $M=600$ GeV (and $c_H = 1$), the correction amounts to $\delta_H = 39\%$, thus leading to a significant enhancement of the partial width.
The effect also scales approximately with $1/M^2$, thus becoming less relevant for larger masses: for instance, for $M = 1$ TeV, we obtain $\delta_H = 15\%$.

This effect being sizeable below 1 TeV and model independent, we decided to include it in the effective Lagrangian in Eq.~(\ref{eq:param}), without introducing a new parameter.

\subsection*{Z: $T \to Z t$}

For the coupling of the $Z$ to the top, the sub-leading coupling is present only in two cases: for $(2,7/6)$ proportional to the leading one times $- 2 \sqrt{\epsilon_t}$, and for $(3,-1/3)$ with factor $2 \sqrt{\epsilon_t}$.
In all other cases, the coupling is absent, and the only effect comes from the top mass in the phase space.
The correction can be written, in all cases, as:
\beq
\delta_Z = \sqrt{\lambda (1,\epsilon_Z, \epsilon_t)}\frac{ ((1-\epsilon_t)^2 +\epsilon_Z- 2 \epsilon_Z^2 + \epsilon_z \epsilon_t)(1+ c_Z^2 \epsilon_t) -12 c_Z \epsilon_Z \epsilon_t }{ \left( 1-3 \epsilon_Z^2 + 2 \epsilon_Z^3\right)} -1 \sim (c_Z^2-3) \epsilon_t\,,
\eeq
where $\epsilon_Z = m_Z^2/M^2$ and $c_Z$ parametrises the sub-leading coupling: it is $c_Z = -2$ for $(2,7/6)$, $c_Z = 2$ for $(3,-1/3)$ and $c_Z = 0$ in all other cases. 
In this case we obtain much smaller corrections than in the Higgs case: for $c_Z = 0$, we obtain $\delta_Z = -23\%$ for $M=600$ GeV, scaling down to $\delta_Z = -9\%$ at a TeV.
In the other two cases, even smaller corrections are obtained, as for $c_Z = 2$ ($c_Z = -2$) we obtain $\delta_Z = 1.5\%$ ($7\%$).
This correction is highly model-dependent, but also numerically small, so we decided to neglect it for the time being.

\subsection*{W: $T \to W^+ b$}

In this decay, the only contribution comes from the sub-leading coupling, which is present only for the doublets.
The correction would therefore read
\beq
\delta_W = c_W^2 \epsilon_t\,.
\eeq
For $(2,1/6)_{\lambda_u = \lambda_d}$, we have $c_W = 1$, and the correction amounts to $\delta_W = 8\%$ ($3\%$) for $M = 600$ GeV ($1$ TeV).

For the $(2,1/6)_{\lambda_d=0}$ and $(2,-5/6)$, the leading coupling vanishes, so $\xi_W^T = 0$.
In this case, one can calculate
\beq
\xi_W^T \delta_W = \frac{1-3 \epsilon_W^2 + 2 \epsilon_W^3}{1-\epsilon_H} \epsilon_t \sim \epsilon_t\,,
\eeq
in both cases.
Numerically, we obtain results similar to the other case.

\subsection*{B: $B \to W^- t$}

This is the only channel in the $B$ decays that is sensitive to the top mass, both in the couplings and phase space.
We can again parametrise the sub-leading coupling to be proportional to the leading one with a factor $c_W$.
The correction can be written as:
\beq
\delta_B = \sqrt{\lambda (1,\epsilon_W, \epsilon_t)} \frac{ ((1-\epsilon_t)^2 +\epsilon_W- 2 \epsilon_W^2 + \epsilon_W \epsilon_t)(1+ c_W^2 \epsilon_t) -12 c_W \epsilon_W \epsilon_t }{\left( 1-3 \epsilon_W^2 + 2 \epsilon_W^3\right)} \sim (c_W^2-3) \epsilon_t\,.
\eeq
The value of $c_W$ depends crucially on the representation $B$ belongs to:
\beq
(1,-1/3) \; \mbox{\&}\; (2,-5/6) & c_W = 0 & \Rightarrow \delta_B \sim -23\% \; (-9\%)\,, \nonumber \\ 
(2,-1/6)_{\lambda_d=0} \; \mbox{\&}\; (2,-1/6)_{\lambda_u=\lambda_d} &  c_W = 1 & \Rightarrow \delta_B \sim -18\%\; (-6\%)\,, \nonumber \\ 
(3,-1/3) & c_W = \sqrt{2} & \Rightarrow \delta_B \sim -13\%\; (-4\%)\,; \nonumber 
\eeq
where the values are calculated for $M = 600$ GeV ($1$ TeV).

A special case is offered by the $(3,2/3)$ case: in fact, for this representation, the leading coupling of the $W$ vanishes, $\xi_W^B = 0$, however the sub-leading coupling is non-vanishing.
A straightforward calculation leads to
\beq
\xi_W \delta_B \sim - \epsilon_t \sim -9\%\; (-3\%)\,.
\eeq

In all cases, the corrections are fairly small.

\subsection*{X: $X \to W^+ t$}

Once again this decay mode is sensitive to the top mass both via the phase space and via the sub-leading couplings, which is present in both cases and proportional to the leading one times $c_W \sqrt{\epsilon_t}$.
The correction can be written as:
\beq
\delta_X = \sqrt{\lambda (1,\epsilon_W, \epsilon_t)} \frac{ ((1-\epsilon_t)^2 +\epsilon_W- 2 \epsilon_W^2 + \epsilon_W \epsilon_t)(1+ c_W^2 \epsilon_t) -12 c_W \epsilon_W \epsilon_t }{\left( 1-3 \epsilon_W^2 + 2 \epsilon_W^3\right)} \sim (c_W^2-3) \epsilon_t\,,
\eeq
where $c_W = 1$ for the doublet $(2,7/6)$ and $c_W = 1/\sqrt{2}$ for the triplet $(3,2/3)$.
Numerically, for the doublet we find $\delta_X = -18\%$ ($-6\%$) and for the triplet $\delta_X = -21\%$ ($-8\%$) for $M=600$ GeV ($1$ TeV).

\section{FeynRules implementation}
\label{app:feynrules}

In this Appendix we summarise the implementation of the model-independent
parametrisation of (\ref{eq:param}) in FeynRules~\cite{wwwFeynRules}. As detailed in Section \ref{subsec:param},
we restrict the present analysis to the case of the four generic states $%
X_{5/3}$, $T_{2/3}$, $B_{-1/3}$ and $Y_{-4/3}$ which can decay directly into a pair of standard model particles.
The common procedure for introducing such new states is to define new class members within a
given $SU(2)_{L}$ representation, with appropriate \verb+Indices+ definitions for each particle class. In the present study, we start
from the Standard Model implementation and add them as the new
coloured spin 1/2\ objects: 
\bigskip 

\begin{minipage}[t]{0.5\textwidth}
\begin{verbatim}
F[5] == {
         ClassName -> xq,
         ClassMembers -> {x},
         SelfConjugate -> False,
         Indices -> {Index[Colour]},
         QuantumNumbers -> {Q -> 5/3},
         Mass -> {MX,600},
         PDG -> {6000005},
         },
 
F[7] == {
         ClassName -> bpq,
         ClassMembers -> {bp},
         SelfConjugate -> False,
         Indices -> {Index[Colour]},
         QuantumNumbers -> {Q -> -1/3},
         Mass -> {MBP,600},
         PDG -> {6000007}
         },
 
\end{verbatim}

\end{minipage}\begin{minipage}[t]{0.5\textwidth}
\begin{verbatim}
F[6] == {
         ClassName -> tpq,
         ClassMembers -> {tp},
         SelfConjugate -> False,
         Indices -> {Index[Colour]},
         QuantumNumbers -> {Q -> 2/3},
         Mass -> {MTP,600},
         PDG -> {6000006},
         },
 
F[8] == {
         ClassName -> yq,
         ClassMembers -> {y},
         SelfConjugate -> False,
         Indices -> {Index[Colour]},
         QuantumNumbers -> {Q -> -4/3},
         Mass -> {MY,600},
         PDG -> {6000008}
         }
\end{verbatim}

\end{minipage}

\bigskip 

Here each \verb+ClassName+ defines a specific VL quark class with a given electric
charge. Such definitions are made without any assumptions on the other quantum numbers. Any change in the above PDG codes should not interfere with the
existing assignments. The masses are set to 600\ GeV\ by default, while the total
widths should be systematically evaluated within MadGraph and
given as inputs in the corresponding parameter cards. For an appropriate
evaluation of the $2\rightarrow 2$ processes cross-sections, all light
quarks included in the proton definition are restricted to be massless (5F
scheme). 

In addition to the VL quark masses (and widths, unless they are computed
automatically), the remaining parameters are divided into the
three \verb+External+ classes \verb+KAPPA+, \verb+XI+ and \verb+ZETA+, combined internally in
FeynRules to match the effective couplings defined in Eq. (\ref%
{eq:param}) :

\bigskip 
\begin{verbatim}
(* Block Kappa *)
 
\end{verbatim}

\begin{minipage}[t]{0.5\textwidth}
\begin{verbatim}
KX == {
         ParameterType -> External,
         BlockName -> Kappa,
         ComplexParameter -> False,
         Description -> "Kappa_X parameter"
         },
 
...
\end{verbatim}

\end{minipage}\begin{minipage}[t]{0.5\textwidth}
\begin{verbatim}
KT == {
         ParameterType -> External,
         BlockName -> Kappa,
         ComplexParameter -> False,
         Description -> "Kappa_T parameter"
         },
 
...
\end{verbatim}

\end{minipage}

\bigskip 
\begin{verbatim}
(* Block Xi *)
\end{verbatim}

\begin{minipage}[t]{0.5\textwidth}
\begin{verbatim}
xitpw == {
         ParameterType -> External,
         BlockName -> Xi,
         ComplexParameter -> False,
         Description -> "BR of T in W"
         },
 
...
\end{verbatim}

\end{minipage}\begin{minipage}[t]{0.5\textwidth}
\begin{verbatim}
xitpz == {
         ParameterType -> External,
         BlockName -> Xi,
         ComplexParameter -> False,
         Description -> "BR ratio of T in Z"
         },
 
...
\end{verbatim}

\end{minipage}

\bigskip 
\begin{verbatim}
(* Block Zeta *)
\end{verbatim}

\begin{minipage}[t]{0.5\textwidth}
\begin{verbatim}
zetaTuL == {
         ParameterType -> External,
         BlockName -> Zeta,
         ComplexParameter -> False,
         Description -> "T-u mixing (LH)"
         },
\end{verbatim}

\end{minipage}\begin{minipage}[t]{0.5\textwidth}
\begin{verbatim}
zetaTcL == {
         ParameterType -> External,
         BlockName -> Zeta,
         ComplexParameter -> False,
         Description -> "T-c mixing (LH)"
         },
\end{verbatim}

\end{minipage}
\begin{verbatim}
...
 
\end{verbatim}

Having defined the new quark fields, the interactions given in Eq. (\ref{eq:param}) are
directly added to the Standard Model Lagrangian, together with the corresponding
kinetic and mass terms for all VL quark species. The electromagnetic and strong currents are implemented as
well. Model extensions for the effective $Qqg(g)$ chromomagnetic strong interaction (\ref{eq:anomalous}) and the off-diagonal charged-current transitions (\ref{eq:offdiagonallagrangian}) can be
loaded together with the generic FeynRules model. All the model files are in UFO format and support the unitary gauge (\verb+Set+ \verb+FeynmanGauge+ \verb+=+ \verb+False+). We comment that, depending on the representation they belong to, the new VL quarks can induce sizeable corrections to the Standard Model quark mixings, which should in principle be included.
However, for the LHC phenomenology, they are irrelevant so that we restrict the implementation to the model independent couplings involving a single VL quark, without limiting the utility of such implementation. 

As a first step of the validation procedure, all the tree-level decay rates have been checked to be in agreement with the analytical formulae (\ref{eq:GammaW})-(\ref{eq:GammaH}) for various benchmark points. Furthermore, the particles decay widths and branching ratios have been calculated with BRIDGE \cite{Meade:2007js}, and successfully compared to the analytical formulae. As a second step, a comparison of the leading order coefficients $\bar{\sigma}^{Qt}$ and $\bar{\sigma}^{QV}$ has been performed between MadGraph5 \cite{Alwall:2011uj} and an independent model implementation of Eq. (\ref{eq:param}) in CalcHEP 3.4 \cite{Belyaev:2012qa}, for similar parameter choices. 
Although deviations up to 10\% can be obtained for particular cases of $\bar{\sigma}^{Qq}$ between MadGraph5 and CalcHEP, 
the $Qq$ channels rates obtained from the UFO\ output have been verified to be consistent with \cite{Atre:2011ae}.
Considering the case of exclusive couplings to the first generation, $\zeta _{1}=1$, cross-sections agree at the percent\ level  when comparing $\sigma
(pp\rightarrow Qq)$ versus $\sigma (pp\rightarrow \bar{Q}q)$ for all
four VL\ quark types and the two benchmark points given in the reference. Finally, we have checked that the MadGraph cross-sections for pair and electroweak single production of top partners at the LHC at $\sqrt{s}=7,8$ and 14\ TeV\ match the leading order predictions for the top quark in the Standard Model, when adjusting the mass and width values. Overall consistency at the \% level is obtained. 

In Tabs. \ref{tab:sigmasparticles600}-\ref{tab:sigmasantiparticles1000} we provide the expansion coefficients for the single production cross sections of all the VLQ species discussed in this paper. The cross sections have been computed at different LHC energies for particles with mass $M=600$, $800$, and $1000$ GeV. 
In Tab. \ref{tab:sigmasAnomalous} we report the contribution to the single $T$ and $B$ production of the chromo-magnetic operator. Finally, in Tabs. \ref{tab:samesignQQ}-\ref{tab:offdiagonalQQ} we list the coefficients for EW pair production of $QQ^{(\prime)}$ pairs.

\begin{table}[tb]
\begin{eqnarray*}
\begin{array}{c|lll|lll|lll}
\toprule
                        & \multicolumn{3}{c|}{7 \text{ TeV}}                 & \multicolumn{3}{c|}{8 \text{ TeV}}                 & \multicolumn{3}{c}{14 \text{ TeV}}                 \\
                        & i=1 & i=2 & i=3 & i=1 & i=2 & i=3 & i=1 & i=2 & i=3 \\
\midrule
\bar\sigma_{Wi}^{T \bar t}     & 893        & 68.4       & 20.7       & 1441         & 123        & 39.1       & 7580         & 985        & 373        \\
\bar\sigma_{Zi}^{T \bar t}     & -            & -            & 4.22      & -            & -            & 6.28      & -            & -            & 2.47       \\
\bar\sigma_{Wi}^{B t}      & 2314         & 34.5       & -            & 3605         & 64.5       & -            & 16700         & 588        & -            \\
\bar\sigma_{Wi}^{B \bar t} & -            & -            & 2.04      & -            & -            & 3.14      & -            & -            & 13.8       \\
\bar\sigma_{Wi}^{X\bar t}  & 2277         & 33.9       & 7.01      & 3546         & 63.2       & 10.0       & 16640         & 578        & 33.6       \\
\bar\sigma_{Wi}^{Y t}      & 936        & 71.2       & 22.3       & 1507         & 128        & 42.7       & 7911         & 1021         & 405        \\

\midrule
\bar\sigma_{Wi}^{T j}          & 34150         & 4943         & 1906         & 45420 & 7316         & 2957         & 125000 & 29400         & 13970         \\
\bar\sigma_{Zi}^{T j}          & 48000 & 1770         & -            & 63200         & 2760         & -            & 171000          & 13400         & -            \\
\bar\sigma_{Wi}^{B j}      & 39500         & 1140         & -            & 53000         & 2090         & -            & 152000          & 10800         & -            \\
\bar\sigma_{Zi}^{B j}      & 22500         & 3030         & 1130         & 30400         & 4550         & 1790         & 91000         & 19600         & 9080         \\
\bar\sigma_{Wi}^{X j}      & 72900         & 2950         & -            & 94000         & 4520         & -            & 232000          & 20400         & -            \\
\bar\sigma_{Wi}^{Y j}      & 18600         & 2290         & 831        & 25600         & 3510         & 1340         & 80500         & 16200         & 7250         \\

\midrule
\bar\sigma_{Wi}^{T W}          & 1300        & 106        & 32.9       & 2070         & 187        & 60.9       & 10700         & 1420         & 545        \\
\bar\sigma_{Wi}^{B W}      & 3270         & 53.5       & -            & 5040         & 97.9       & -            & 23400         & 840        & -            \\
\bar\sigma_{Wi}^{X W}      & 3270         & 53.5       & -            & 5040         & 97.9       & -            & 23400         & 840        & -            \\
\bar\sigma_{Wi}^{Y W}      & 1300         & 106        & 33.0       & 2070         & 187        & 60.9       & 10700         & 1420         & 545        \\

\bar\sigma_{Zi}^{T Z}          & 3370         & 55.0       & -            & 5200         & 101        & -            & 24200         & 869        & -            \\
\bar\sigma_{Zi}^{B Z}      & 1340         & 109        & 33.9       & 2130         & 193        & 62.6       & 11100         & 1470         & 563        \\
\bar\sigma_{Hi}^{T H}          & 2460         & 34.5       & -            & 3610         & 64.5       & -            & 16900         & 588        & -            \\

\bar\sigma_{Hi}^{B H}      & 965        & 74.1       & 22.5       & 1560         & 133        & 42.4       & 8560         & 1090         & 409        \\
\bottomrule
\end{array}
\end{eqnarray*}
\caption{Coefficients (in fb) for single production of VL quarks with mass $M=600$ GeV. $\bar\sigma_C^{A B}$ is the 
coefficient in the expansion, corresponding to the production of a VL quark $A$ in association with a particle $B$ due to the exchange of $C$, where $j$ labels a jet (including the $b$ quark). Bottom quarks have been included among proton components and as final states 
in $Q j$ processes. Contributions of interference terms are neglected as we checked they give a negligible effect.}
\label{tab:sigmasparticles600}
\end{table}

\begin{table}[tb]
\begin{eqnarray*}
\begin{array}{c|lll|lll|lll}
\toprule
                             & \multicolumn{3}{c|}{7 \text{ TeV}}                 & \multicolumn{3}{c|}{8 \text{ TeV}}                 & \multicolumn{3}{c}{14 \text{ TeV}}                 \\
                        & i=1 & i=2 & i=3 & i=1 & i=2 & i=3 & i=1 & i=2 & i=3 \\
\midrule
\bar\sigma_{Wi}^{\bar T t}      & 138 & 68.7       & 20.7       & 244        & 124        & 39.1       & 1800         & 992        & 373        \\
\bar\sigma_{Zi}^{\bar T t}      & -            & -            & 4.21      & -            & -            & 6.28      & -            & -            & 24.7       \\
\bar\sigma_{Wi}^{\bar B \bar t} & 105        & 34.4       & -            & 186        & 64.5       & -            & 1410         & 590        & -            \\
\bar\sigma_{Wi}^{\bar B t}      & -            & -            & 6.36      & -            & -            & 9.21      & -            & -            & 32.3       \\
\bar\sigma_{Wi}^{\bar X t}      & 103        & 34.0       & 2.31      & 184        & 63.8       & 3.51      & 1390         & 581        & 14.9       \\
\bar\sigma_{Wi}^{\bar Y\bar t}  & 143        & 71.2       & 22.5       & 254        & 128        & 42.6       & 1860         & 1020         & 405        \\
\midrule
\bar\sigma_{Wi}^{\bar T j}      & 4010         & 2160         & 783        & 6040 & 3330         & 1270         & 25600         & 15500         & 6950         \\
\bar\sigma_{Zi}^{\bar T j}      & 4010         & 1680         & -            & 5990 & 2620         & -            & 25300         & 12900         & -            \\
\bar\sigma_{Wi}^{\bar B j}      & 6510 & 2780         & -            & 9590         & 4260         & -            & 37400         & 19500 & -            \\
\bar\sigma_{Zi}^{\bar B j}      & 5240 & 2870 & 1080         & 7720         & 4330         & 1700         & 30500         & 19000         & 8720         \\
\bar\sigma_{Wi}^{\bar X j}      & 3030         & 1240         & -            & 4640 & 1970         & -            & 20900         & 10400         & -            \\
\bar\sigma_{Wi}^{\bar Y j}      & 8380         & 4660         & 1780         & 12100         & 6930         & 2770         & 44400         & 28300         & 13300         \\
\midrule
\bar\sigma_{Wi}^{\bar T W}      & 210        & 106        & 33.0       & 365        & 187        & 60.9       & 2570         & 1420         & 545        \\
\bar\sigma_{Wi}^{\bar B W}      & 158        & 53.7       & -            & 275        & 98.1       & -            & 1990         & 848        & -            \\
\bar\sigma_{Wi}^{\bar X W}      & 158        & 53.7       & -            & 275        & 98.1       & -            & 2000         & 846        & -            \\
\bar\sigma_{Wi}^{\bar Y W}      & 210        & 106        & 33.0       & 365        & 187        & 60.9       & 2570         & 1420         & 545        \\
\bar\sigma_{Zi}^{\bar T Z}      & 163        & 55.2       & -            & 283        & 101        & -            & 2060         & 870        & -            \\
\bar\sigma_{Zi}^{\bar B Z}      & 216        & 109        & 33.9       & 376        & 193        & 62.6       & 2650         & 1470         & 563        \\
\bar\sigma_{Hi}^{\bar T H}      & 111 & 36.8       & -            & 198        & 68.8       & -            & 1540         & 637        & -            \\
\bar\sigma_{Hi}^{\bar B H}      & 148 & 73.9       & 22.5       & 263        & 134        & 42.44       & 1990         & 1090 & 409 \\
\bottomrule
\end{array}
\end{eqnarray*}
\caption{The same as Tab.\ref{tab:sigmasparticles600} for VL antiquarks.}
\label{tab:sigmasantiparticles600}
\end{table}


\begin{table}[tb]
\begin{eqnarray*}
\begin{array}{c|lll|lll|lll}
\toprule
                        & \multicolumn{3}{c|}{7 \text{ TeV}}                 & \multicolumn{3}{c|}{8 \text{ TeV}}                 & \multicolumn{3}{c}{14 \text{ TeV}}                 \\
                        & i=1 & i=2 & i=3 & i=1 & i=2 & i=3 & i=1 & i=2 & i=3 \\
\midrule
\bar\sigma_{Wi}^{T \bar t}     & 398        & 24.8       & 7.07       & 692         & 49.0        & 14.6       & 4520         & 508        & 183        \\
\bar\sigma_{Zi}^{T \bar t}     & -            & -            & 1.06      & -            & -            & 1.69      & -            & -            & 8.28       \\
\bar\sigma_{Wi}^{B t}      & 1090         & 11.8       & -            & 1820         & 24.1       & -            & 10400         & 289        & -            \\
\bar\sigma_{Wi}^{B \bar t} & -            & -            & 0.485      & -            & -            & 0.803      & -            & -            & 4.42       \\
\bar\sigma_{Wi}^{X\bar t}  & 1080         & 11.7       & 1.90      & 1810         & 23.9       & 2.92       & 10300         & 287        & 12.1       \\
\bar\sigma_{Wi}^{Y t}      & 418        & 25.9       & 7.59       & 724         & 50.9        & 15.7       & 4710         & 525         & 196        \\

\midrule
\bar\sigma_{Wi}^{T j}          & 16910 & 1915 & 674         & 23750 & 3040         & 1130         & 75400 & 14900         & 6670         \\
\bar\sigma_{Zi}^{T j}          & 24400 & 618 & -            & 33700         & 1040         & -            & 104000          & 6290         & -            \\
\bar\sigma_{Wi}^{B j}      & 19200         & 436         & -            & 27200         & 748         & -            & 90500          & 4940         & -            \\
\bar\sigma_{Zi}^{B j}      & 10680         & 1130         & 392         & 15300         & 1830         & 665         & 53000         & 9700         & 4230         \\
\bar\sigma_{Wi}^{X j}      & 38600         & 1060         & -            & 52100         & 1740         & -            & 147000          & 9880         & -            \\
\bar\sigma_{Wi}^{Y j}      & 8400         & 814         & 272        & 12300         & 1340         & 473         & 45400         & 7730         & 3270         \\

\midrule
\bar\sigma_{Wi}^{T W}          & 363 & 23.2 & 6.70       & 631         & 45.3 & 13.7       & 4260         & 470         & 170        \\
\bar\sigma_{Wi}^{B W}      & 984         & 11.0       & -            & 1650         & 22.2       & -            & 9850         & 264        & -            \\
\bar\sigma_{Wi}^{X W}      & 984         & 11.0       & -            & 1650         & 22.2       & -            & 9850         & 264        & -            \\
\bar\sigma_{Wi}^{Y W}      & 363         & 23.2        & 6.70       & 631         & 45.3        & 13.7       & 4260         & 470         & 170        \\

\bar\sigma_{Zi}^{T Z}          & 1010         & 11.2       & -            & 1690         & 22.8        & -            & 10100         & 270        & -            \\
\bar\sigma_{Zi}^{B Z}      & 372         & 23.7        & 6.85       & 646         & 46.3        & 14.0       & 4360         & 480         & 174        \\
\bar\sigma_{Hi}^{T H}          & 794         & 8.18       & -            & 1350         & 16.9       & -            & 8460 & 216        & -            \\

\bar\sigma_{Hi}^{B H}      & 289        & 17.6       & 4.98       & 511         & 35.0        & 10.4       & 3630         & 388         & 138        \\
\bottomrule
\end{array}
\end{eqnarray*}
\caption{The same as Tab.\ref{tab:sigmasparticles600} for $M=800$ GeV.}
\label{tab:sigmasparticles800}
\end{table}

\begin{table}[tb]
\begin{eqnarray*}
\begin{array}{c|lll|lll|lll}
\toprule
                             & \multicolumn{3}{c|}{7 \text{ TeV}}                 & \multicolumn{3}{c|}{8 \text{ TeV}}                 & \multicolumn{3}{c}{14 \text{ TeV}}                 \\
                        & i=1 & i=2 & i=3 & i=1 & i=2 & i=3 & i=1 & i=2 & i=3 \\
\midrule
\bar\sigma_{Wi}^{\bar T t}      & 50.7 & 25.0       & 7.11       & 99.0        & 49.2        & 14.7       & 953         & 509        & 183        \\
\bar\sigma_{Zi}^{\bar T t}      & -            & -            & 1.05      & -            & -            & 1.69      & -            & -            & 8.28       \\
\bar\sigma_{Wi}^{\bar B \bar t} & 39.0        & 11.8       & -            & 75.2        & 24.2       & -            & 734         & 290        & -            \\
\bar\sigma_{Wi}^{\bar B t}      & -            & -            & 1.68      & -            & -            & 2.62      & -            & -            & 11.3       \\
\bar\sigma_{Wi}^{\bar X t}      & 38.8        & 11.7       & 0.563      & 74.7        & 24.0       & 0.920      & 727         & 287        & 4.86       \\
\bar\sigma_{Wi}^{\bar Y\bar t}  & 52.6        & 25.8       & 7.56       & 103        & 50.9        & 15.7       & 988         & 525         & 196        \\
\midrule
\bar\sigma_{Wi}^{\bar T j}      & 1510 & 778         & 259        & 2440 & 1290         & 452         & 12900         & 7500         & 3160         \\
\bar\sigma_{Zi}^{\bar T j}      & 1550         & 591         & -            & 2480 & 994         & -            & 12800         & 6120         & -            \\
\bar\sigma_{Wi}^{\bar B j}      & 2610 & 1010         & -            & 4100         & 1670         & -            & 19500         & 9510 & -            \\
\bar\sigma_{Zi}^{\bar B j}      & 2080 & 1090 & 374         & 3290         & 1760         & 636         & 15900         & 9440         & 4100         \\
\bar\sigma_{Wi}^{\bar X j}      & 1120         & 414         & -            & 1830 & 714         & -            & 10200         & 4770         & -            \\
\bar\sigma_{Wi}^{\bar Y j}      & 3450         & 1820         & 638         & 5350 & 2910         & 1070         & 23900         & 14400         & 6420         \\
\midrule
\bar\sigma_{Wi}^{\bar T W}      & 46.9        & 23.2        & 6.70       & 90.8        & 45.3        & 13.7       & 882         & 471         & 170        \\
\bar\sigma_{Wi}^{\bar B W}      & 35.9        & 11.0       & -            & 68.8        & 22.3       & -            & 673 & 266        & -            \\
\bar\sigma_{Wi}^{\bar X W}      & 35.9               & 11.0       & -            & 68.8        & 22.3       & -            & 673 & 266 & -            \\
\bar\sigma_{Wi}^{\bar Y W}      & 46.9 & 23.2 & 6.70       & 90.8        & 45.3        & 13.7       & 882         & 471         & 170        \\
\bar\sigma_{Zi}^{\bar T Z}      & 36.7 & 11.3       & -            & 70.4        & 22.8        & -            & 689         & 272 & -            \\
\bar\sigma_{Zi}^{\bar B Z}      & 48.1        & 23.7        & 6.85       & 93.0        & 46.4        & 14.0       & 902         & 482         & 174        \\
\bar\sigma_{Hi}^{\bar T H}      & 27.4 & 8.23       & -            & 53.6        & 17.0       & -            & 556         & 215        & -            \\
\bar\sigma_{Hi}^{\bar B H}      & 35.6 & 17.6       & 4.98       & 70.5        & 35.0        & 10.4       & 731         & 387 & 138 \\
\bottomrule
\end{array}
\end{eqnarray*}
\caption{The same as Tab.\ref{tab:sigmasantiparticles600} for $M=800$ GeV.}
\label{tab:sigmasantiparticles800}
\end{table}

\begin{table}[tb]
\begin{eqnarray*}
\begin{array}{c|lll|lll|lll}
\toprule
                        & \multicolumn{3}{c|}{7 \text{ TeV}}                 & \multicolumn{3}{c|}{8 \text{ TeV}}                 & \multicolumn{3}{c}{14 \text{ TeV}}                 \\
                        & i=1 & i=2 & i=3 & i=1 & i=2 & i=3 & i=1 & i=2 & i=3 \\
\midrule
\bar\sigma_{Wi}^{T \bar t}     & 183        & 9.44       & 2.56       & 343         & 20.4        & 5.80       & 2800         & 275        & 95.1        \\
\bar\sigma_{Zi}^{T \bar t}     & -            & -            & 0.302      & -            & -            & 0.526      & -            & -            & 3.23       \\
\bar\sigma_{Wi}^{B t}      & 534         & 4.27       & -            & 955         & 9.56       & -            & 6700 & 150 & -            \\
\bar\sigma_{Wi}^{B \bar t} & -            & -            & 0.133      & -            & -            & 0.239      & -            & -            & 1.66       \\
\bar\sigma_{Wi}^{X\bar t}  & 528         & 4.23       & 0.577      & 947         & 9.50       & 0.965       & 6650         & 150        & 4.99       \\
\bar\sigma_{Wi}^{Y t}      & 192        & 9.83       & 2.72       & 360         & 21.2        & 6.17       & 2920         & 285         & 101        \\

\midrule
\bar\sigma_{Wi}^{T j}          & 8830 & 800 & 261         & 13100 & 1370         & 470         & 48700 & 8250         & 3490         \\
\bar\sigma_{Zi}^{T j}          & 13100 & 236 & -            & 19100         & 428         & -            & 68000          & 3260         & -            \\
\bar\sigma_{Wi}^{B j}      & 9850         & 158         & -            & 14800         & 296         & -            & 57600          & 2470         & -            \\
\bar\sigma_{Zi}^{B j}      & 5380         & 460         & 148         & 8160         & 800         & 272         & 33200         & 5250         & 2170         \\
\bar\sigma_{Wi}^{X j}      & 21400         & 414         & -            & 30500 & 739 & -            & 98900         & 5220         & -            \\
\bar\sigma_{Wi}^{Y j}      & 4040         & 316         & 97.8        & 6280         & 566         & 185         & 27700         & 4070         & 1630         \\

\midrule
\bar\sigma_{Wi}^{T W}          & 116 & 5.94 & 1.63       & 220         & 12.9 & 3.70       & 1940         & 183         & 62.6        \\
\bar\sigma_{Wi}^{B W}      & 337         & 2.68       & -            & 613         & 6.03       & -            & 4700         & 98.2        & -            \\
\bar\sigma_{Wi}^{X W}      & 337         & 2.68       & -            & 613         & 6.03       & -            & 4700         & 98.2        & -            \\
\bar\sigma_{Wi}^{Y W}      & 116         & 5.94        & 1.63       & 220         & 12.9        & 3.70             & 1940         & 183         & 62.6               \\

\bar\sigma_{Zi}^{T Z}          & 343         & 2.73       & -            & 625 & 6.16        & -            & 4780         & 99.8        & -            \\
\bar\sigma_{Zi}^{B Z}      & 118         & 6.06        & 1.66       & 225         & 13.2        & 3.76       & 1970         & 186         & 63.7        \\
\bar\sigma_{Hi}^{T H}          & 283 & 2.10       & -            & 524         & 4.84       & -            & 4190 & 83.6        & -            \\

\bar\sigma_{Hi}^{B H}      & 96.2        & 4.73       & 1.28       & 186         & 10.5        & 2.96       & 1720         & 158         & 53.2        \\
\bottomrule
\end{array}
\end{eqnarray*}
\caption{The same as Tab.\ref{tab:sigmasparticles600} for $M=1000$ GeV.}
\label{tab:sigmasparticles1000}
\end{table}

\begin{table}[tb]
\begin{eqnarray*}
\begin{array}{c|lll|lll|lll}
\toprule
                             & \multicolumn{3}{c|}{7 \text{ TeV}}                 & \multicolumn{3}{c|}{8 \text{ TeV}}                 & \multicolumn{3}{c}{14 \text{ TeV}}                 \\
                        & i=1 & i=2 & i=3 & i=1 & i=2 & i=3 & i=1 & i=2 & i=3 \\
\midrule
\bar\sigma_{Wi}^{\bar T t}      & 19.4 & 9.49       & 2.58       & 41.7        & 20.5        & 5.83       & 529         & 276        & 95.4        \\
\bar\sigma_{Zi}^{\bar T t}      & -            & -            & 0.302      & -            & -            & 0.526      & -            & -            & 3.23       \\
\bar\sigma_{Wi}^{\bar B \bar t} & 15.3        & 4.26       & -            & 32.1        & 9.59       & -            & 402         & 151        & -            \\
\bar\sigma_{Wi}^{\bar B t}      & -            & -            & 0.499      & -            & -            & 0.850      & -            & -            & 4.63       \\
\bar\sigma_{Wi}^{\bar X t}      & 15.3        & 4.26       & 0.158      & 32.1        & 9.55       & 0.280      & 400         & 150        & 1.85       \\
\bar\sigma_{Wi}^{\bar Y\bar t}  & 20.1        & 9.81       & 2.71       & 43.3        & 21.2        & 6.14       & 549         & 285         & 101        \\
\midrule
\bar\sigma_{Wi}^{\bar T j}      & 606 & 303         & 93.8        & 1070 & 544         & 178         & 7040         & 3950         & 1580         \\
\bar\sigma_{Zi}^{\bar T j}      & 654         & 227         & -            & 1120 & 414         & -            & 7030         & 3180         & -            \\
\bar\sigma_{Wi}^{\bar B j}      & 1130 & 398         & -            & 1900         & 710 & -            & 11000         & 5080 & -            \\
\bar\sigma_{Zi}^{\bar B j}      & 881 & 445 & 142         & 1500         & 775         & 262         & 8930         & 5120         & 2110         \\
\bar\sigma_{Wi}^{\bar X j}      & 450         & 152         & -            & 793 & 285         & -            & 5470         & 2400         & -            \\
\bar\sigma_{Wi}^{\bar Y j}      & 1500 & 769 & 249         & 2530         & 1320         & 450         & 13800         & 8020         & 3380         \\
\midrule
\bar\sigma_{Wi}^{\bar T W}      & 12.2        & 5.98        & 1.63       & 26.3        & 13.0        & 3.70       & 351         & 183         & 62.6        \\
\bar\sigma_{Wi}^{\bar B W}      & 9.63        & 2.69       & -            & 20.3        & 6.04       & -            & 266         & 98.7        & -            \\
\bar\sigma_{Wi}^{\bar X W}      & 9.63        & 2.69       & -            & 20.3        & 6.04       & -            & 266         & 98.7        & -            \\
\bar\sigma_{Wi}^{\bar Y W}      & 12.2        & 5.98        & 1.63       & 26.3        & 13.0        & 3.70       & 351         & 183         & 62.6        \\
\bar\sigma_{Zi}^{\bar T Z}      & 9.80        & 2.74       & -            & 20.7        & 6.15        & -            & 271         & 101        & -            \\
\bar\sigma_{Zi}^{\bar B Z}      & 12.4        & 6.09        & 1.66       & 26.7        & 13.2        & 3.76       & 358         & 187         & 63.7        \\
\bar\sigma_{Hi}^{\bar T H}      & 7.72 & 2.11       & -            & 16.6        & 4.84       & -            & 230         & 84.0        & -            \\
\bar\sigma_{Hi}^{\bar B H}      & 9.68 & 4.75       & 1.28       & 21.4        & 10.5        & 2.96       & 304         & 157 & 53.2 \\
\bottomrule
\end{array}
\end{eqnarray*}
\caption{The same as Tab.\ref{tab:sigmasantiparticles600} for $M=1000$ GeV.}
\label{tab:sigmasantiparticles1000}
\end{table}


\begin{table}[tb]
\begin{eqnarray*}
\begin{array}{c|ccc|ccc|ccc}
\toprule
& \multicolumn{3}{c|}{7\text{ TeV}} & \multicolumn{3}{c|}{8\text{ TeV}} & 
\multicolumn{3}{c}{14\text{ TeV}} \\ 
& f_{L}^{1}=1 & f_{L}^{2}=1 & f_{L}^{3}=1 & f_{L}^{1}=1 & f_{L}^{2}=1 & 
f_{L}^{3}=1 & f_{L}^{1}=1 & f_{L}^{2}=1 & f_{L}^{3}=1 \\ 
\midrule 
\sigma_{qg}^{T} & 2.96 & 0.0957 & - & 4.05 & 0.152 & - & 
12.8 & 0.823 & - \\ 
\sigma_{\bar{q}g}^{\bar{T}} & 0.237 & 0.0957 & - & 0.361 & 0.152
& - & 1.68 & 0.822 & - \\ 
\sigma_{qg}^{B} & 1.34 & 0.167 & 0.0610 & 1.88 & 0.257 & 
0.0981 & 6.47 & 1.24 & 0.552 \\ 
\sigma_{\bar{q}g}^{\bar{B}} & 0.309 & 0.166 & 0.0610 & 0.465 & 
0.257 & 0.0981 & 2.06 & 1.24 & 0.552        \\
\bottomrule
\end{array}%
\end{eqnarray*}
\caption{Cross sections (in fb) obtained from the FeynRules
implementation of Eq. (\ref{eq:anomalous}) for $q/\bar{q}$ $g\rightarrow Q/\bar{Q}$ mono production of $T$ and $B$ heavy partners for $%
M=600$ GeV with the choice $\kappa _{g}=1$ and $\Lambda =10$ TeV.\ Different parameter values require to re-scale the numbers by a factor $10^{4}\times \kappa _{g}^{2}/\Lambda ^{4}$ where $\Lambda $ is in TeV. Bottom
quarks are included as initial states. }
\label{tab:sigmasAnomalous}
\end{table}

\begin{table}[tb]
\begin{center}
\begin{tabular}{c|ccc|ccc}
\toprule
& $\zeta_1\zeta_1$ & $\zeta_2\zeta_2$ & $\zeta_3\zeta_3$ & $\zeta_1\zeta_2$ & $\zeta_1\zeta_3$ & $\zeta_2\zeta_3$\\
\midrule
\midrule
\multicolumn{7}{c}{$M=600$ GeV}\\
\midrule
\midrule
$\bar\sigma_Z^{TT}$     &  86503 &  22 &   - & 2224 &    - &   - \\
$\bar\sigma_H^{TT}$     &  80599 &  21 &   - & 2135 &    - &   - \\
\midrule
$\bar\sigma_Z^{BB}$     &  16870 & 101 &   8 & 1597 &  502 &  28 \\
$\bar\sigma_H^{BB}$     &  15683 &  98 &   7 & 1513 &  485 &  27 \\
\midrule
\midrule
\multicolumn{7}{c}{$M=800$ GeV}\\
\midrule
\midrule
$\bar\sigma_Z^{TT}$     &  66857 &   5 &   - &  959 &    - &   - \\
$\bar\sigma_H^{TT}$     &  64077 &   5 &   - &  929 &    - &   - \\
\midrule
$\bar\sigma_Z^{BB}$     &  10395 &  30 &   2 &  675 &  189 &  8 \\
$\bar\sigma_H^{BB}$     &  10092 &  30 &   2 &  659 &  184 &  7 \\
\midrule
\midrule
\multicolumn{7}{c}{$M=1000$ GeV}\\
\midrule
\midrule
$\bar\sigma_Z^{TT}$     &  45858 &   1 &   - &  380 &    - &   - \\
$\bar\sigma_H^{TT}$     &  44920 &   1 &   - &  383 &    - &   - \\
\midrule
$\bar\sigma_Z^{BB}$     &   5701 &   9 & 0.5 &  265 &   68 &   2 \\
$\bar\sigma_H^{BB}$     &   5664 &   9 & 0.5 &  256 &   67 &   2 \\
\bottomrule
\end{tabular}
\caption{Coefficients (in fb) for pair production of same-sign $TT$ and $BB$ for different values of VLQ masses and at an energy of 8 TeV. Bottom quarks have been included among proton components. Contributions of interference terms are neglected as we checked they give a negligible effect.}
\label{tab:samesignQQ}
\end{center}
\end{table}

\begin{table}[tb]
\begin{center}
\begin{tabular}{c|ccc|ccc|ccc}
\toprule
& $\zeta_u\zeta_d$ & $\zeta_u\zeta_s$ & $\zeta_u\zeta_b$ & $\zeta_c\zeta_d$ & $\zeta_c\zeta_s$ & $\zeta_c\zeta_b$ & $\zeta_t\zeta_d$ & $\zeta_t\zeta_s$ & $\zeta_t\zeta_b$\\
\midrule
\midrule
\multicolumn{10}{c}{$M=600$ GeV}\\
\midrule
\midrule
$\bar\sigma_W^{XB}$     & 166903 &  4101 &    - & 4070 &  42 &  - &    - &  - &  - \\
\midrule
$\bar\sigma_W^{TB}$     &  70742 &  1491 &    - & 7672 &  87 &  - & 2550 & 24 &  - \\
$\bar\sigma_Z^{TB}$     &  73275 &  7882 & 2616 & 1530 &  89 & 24 &    - &  - &  - \\
$\bar\sigma_H^{TB}$     &  67750 &  7512 & 2515 & 1473 &  87 & 24 &    - &  - &  - \\
\midrule
$\bar\sigma_W^{TY}$     &  32790 &  2889 &  918 & 2900 & 198 & 52 &  918 & 52 & 15 \\
\midrule
\midrule
\multicolumn{10}{c}{$M=800$ GeV}\\
\midrule
\midrule
$\bar\sigma_W^{XB}$     & 131313 &  1795 &    - & 1791 &  11 &  - &    - &  - &  - \\
\midrule
$\bar\sigma_W^{TB}$     &  50430 &   581 &    - & 3737 &  24 &  - & 1073 &  6 &  - \\
$\bar\sigma_Z^{TB}$     &  50962 &  3769 & 1091 &  590 &  25 &  6 &    - &  - &  - \\
$\bar\sigma_H^{TB}$     &  49375 &  3709 & 1086 &  581 &  25 &  6 &    - &  - &  - \\
\midrule
$\bar\sigma_W^{TY}$     &  20546 &  1260 &  354 & 1261 &  60 & 15 &  354 & 15 &  4 \\
\midrule
\midrule
\multicolumn{10}{c}{$M=1000$ GeV}\\
\midrule
\midrule
$\bar\sigma_W^{XB}$     &  89363 &   717 &    - &  725 &   3 &  - &    - &  - &  - \\
\midrule
$\bar\sigma_W^{TB}$     &  31287 &   216 &    - & 1617 &   7 &  - &  447 &  2 &  - \\
$\bar\sigma_Z^{TB}$     &  31552 &  1667 &  439 &  215 &   7 &  2 &    - &  - &  - \\
$\bar\sigma_H^{TB}$     &  31145 &  1664 &  436 &  214 &   7 &  2 &    - &  - &  - \\
\midrule
$\bar\sigma_W^{TY}$     &  11315 &   499 &  128 &  510 &  17 &  4 &  128 &  4 &  1 \\
\bottomrule
\end{tabular}
\caption{Coefficients (in fb) for pair production of $QQ^\prime$ with $Q\neq Q^\prime$ for different values of VLQ masses and at an energy of 8 TeV. Bottom quarks have been included among proton components. Contributions of interference terms are neglected as we checked they give a negligible effect.}
\label{tab:offdiagonalQQ}
\end{center}
\end{table}

\end{document}